\documentclass[pra,aps,twocolumn,showpacs]{revtex4-1}
\usepackage{graphicx}
\usepackage{bm,color}
\usepackage{physics}
\usepackage{amsmath, amssymb}
\usepackage{bm,color}
\usepackage{multirow}
\usepackage{ulem}

\newcommand{\be}{\begin{eqnarray}}
\newcommand{\ee}{\end{eqnarray}}

\newcommand{\vv}{\overrightarrow}
\usepackage{ulem}

\newcommand{\cbin}[1]{{\color{blue}#1}}
\renewcommand{\cbin}[1]{}

\renewcommand{\theequation}{\arabic{equation}}

\begin{document}

\title{
$Z_2\times Z_2$ symmetry and $Z_4$ Berry phase of bosonic ladder
}
\date{\today}
\author{Yoshihito Kuno$^{1}$}
\author{Yasuhiro Hatsugai$^{2}$}

\affiliation{$^1$Graduate School of Engineering Science, Akita University, Akita 010-8502, Japan}
\affiliation{$^2$Department of Physics, University of Tsukuba, Tsukuba, Ibaraki 305-8571, Japan}

\begin{abstract}
Bose gas on a two-leg ladder exhibits an interesting topological phase. 
We show the presence of a bosonic symmetry-protected-topological (SPT) phase protected by $Z_2\times Z_2$ symmetry. 
This symmetry leads to $Z_4$ fractional quantization of $Z_4$ Berry phase, 
that is a topological order parameter to identify the bulk. 
Using the $Z_4$ Berry phase, we have shown that the interacting bosonic system possesses rich topological phases depending on particle density and strength of interaction. 
Based on the bulk-edge correspondence, each edge state of the SPT phases
is discussed in relation to the $Z_4$ Berry phases. 
Especially we have found an intermediate phase that is not adiabatically connected to a simple adiabatic limit, that possesses unconventional edge states, which we have numerically demonstrate by employing the density-matrix-renormalization group algorithm.
\end{abstract}


\maketitle
\section{Introduction}
Synthetic dimension in cold atom system constituted by internal states of atom gives a promising platform 
to simulate and investigate various states of matter \cite{Ozawa2019}. 
An interesting lattice model has already been implemented, such as a two- or three-leg ladder system with an artificial gauge field \cite{Mancini2015,Stuhl2015,Kang2018,Han2019,Kang2020,Han2022}. 
In coldatom experiments, interacting systems can be implemented and its strength can be controllable \cite{Bloch2008}, e.g., 
Feshbach resonance and using dipole-dipole interactions between dipolar atoms \cite{Trefzger2011,Baier2016}. 
In particular, such a system with specific lattice geometry can be used as a quantum simulator to realize rich topological states of matter.

In the community side of the condensed matter theory, symmetry protected topological (SPT) phase is now an attractive state as one of topological states of matter \cite{Pollmann2012}. 
So far, various types of SPT phases have been discovered and also the classification of the SPT phases has been proposed for some groups of the system. The classification for free fermion systems has been explicitly given \cite{Altland1997,Kitaev2009,Ryu2010} (now called ``ten-fold way") and also even for interacting bosonic system. 
This is listed as a catalog of (bosonic) SPT phases by group cohomology \cite{Chen2012,Chen2013}. 
These classification schemes showed possible SPT phases, but demonstration of concrete examples in realistic system is an ongoing problem. 
In this work, we propose a specific concrete example of an interacting bosonic SPT phase of a two-leg ladder system, feasible in real experiments. 

We consider a Bose-Hubbard model with two internal states of atoms, which can be regarded as a two-leg ladder system. 
Its experimental realization may be easier than that of fermionic one since the system's temperature of the fermionic system in an optical lattice is somewhat high and it is still a challenging problem to observe a complete quantum long-range order, such as magnetization \cite{Mazurenko2017}. We assume that the bosonic model on the two-leg ladder includes on-site and vertical link interactions (interactions between two different internal states), and also a hopping dimerization. 
Then, a specific type of the bosonic interacting SPT phase appears, characterized by fractional quantization of $Z_4$ Berry phase. Due to the two-leg ladder geometry, the system has a key symmetry to induce the SPT phase. 

We find that the key symmetry is $Z_2\times Z_2$ type, which consists of two types of reflection symmetry combined with time-reversal (complex conjugation). 
In this work, we show that $Z_N$ Berry phase \cite{Hatsugai2006,Hirano2008,Maruyama2009,Hatsugai2011,Chepiga2013,Kariyado2015,Maruyama2018,Kariyado2018,Kawarabayashi2019,Mizoguchi2019,Araki2020,KY_2020,Otsuka2021,Bunney2022,HK2023} can be used as a topological order parameter to characterize the bosonic SPT phase. We analytically show that the $Z_4$ Berry phase is fractionally quantized by the presence of the $Z^{*}_2\times Z^{*}_2$ equivalence, as $\gamma=2\pi n/4 \ \text{mod}\, 2\pi$ ($n=1,2,3,4$).
The quantization is protected as far as the gap is opened even under a local twist with the symmetry constraints. 
If the state is adiabatically connected to a set of simple local
clusters (plaquette), we may expect the ground state is short-range entangled and topological properties are reduced to those of the simple one. 
In this work, we numerically demonstrate the presence of the interacting bosonic SPT phases characterized by the fractional quantization of $Z_4$ Berry phase coming from $Z^{*}_2\times Z^{*}_2$ equivalence. Due to the Bose Hubbard nature, we find rich SPT phases depending on mean density and strength of interaction. 
Furthermore, reduction of the equivalence $Z_2^{*}\times Z_2^{*}\to Z^{*}_2$ is demonstrated by the quantized Berry phases of $Z_4\to Z_2$.
Also an eccentric intermediate phase, 
that is not the adiabatically connected to the simple clusters, is
demonstrated.

In relation to the Berry phase of the bulk SPT phase, we numerically discuss edge states based on the bulk-edge correspondence by using the density matrix renormalization algorithm (DMRG). 
We shall show some case studies. 
Depending on the type of edges, mean density and chemical potential, the density profile around the edges significantly varies,
that indicate appearance of unconventional edge states. 
In particular, we numerically find the clear edge density profile, which can be easily identified by higher or lower density than that of the bulk. The appearance of the edge density profile is favorable for the experimental detection of the bosonic SPT phases.

The rest of this paper is organized as follows.
In Sec.~II, we introduce the target model.
In Sec.~III, we analytically show the quantization of the $Z_4$ Berry phase by $Z^{*}_2\times Z^{*}_2$ equivalence.
In Sec.~IV, we numerically investigate the presence of the SPT phases in the model, and, in Sec.~V, turn into the study of the system with an open boundary case and numerically investigate the presence of the edge state corresponding to the SPT phase in the bulk. 
We observe the presence of the edge state where the bulk of the system is in $Z_2\times Z_2$ SPT phase.
Section VI is devoted to discussion and conclusion.

\section{Model}
\begin{figure}[t]
\begin{center} 
\includegraphics[width=9.0cm]{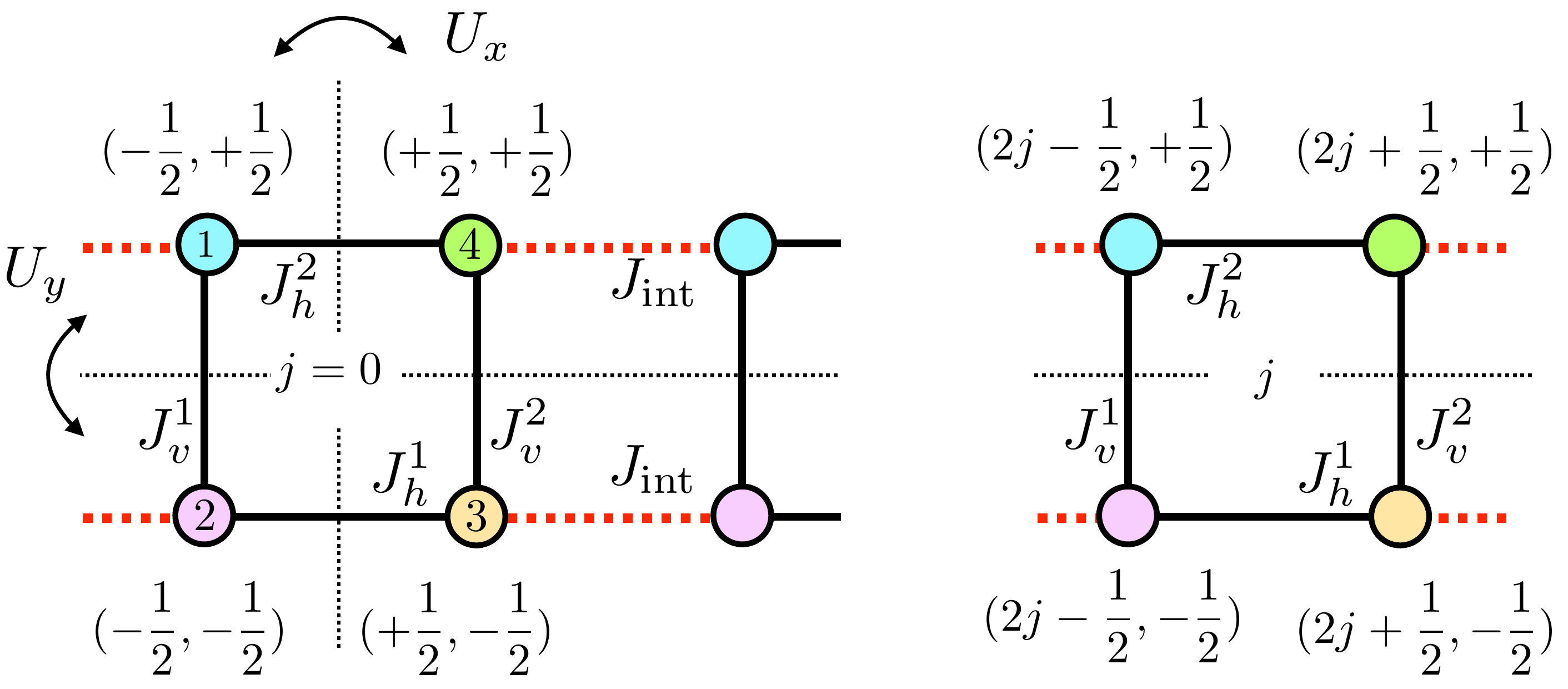}    
\end{center} 
\caption{Labeling of the sites and the lattice structure of the model with periodic boundary condition. Symmetry axes associated with the reflection $U_x$ and $U_y$ are also shown. Note that the open system by cutting the link at the origin ($J_h=0$ at the links $(-\frac 1 2,\pm \frac 1 2  )-(+\frac 1 2,\pm \frac 1 2  )$) also respects these symmetries. It guarantees the bulk-edge correspondence associated with $Z_4$ Berry phase.}
\label{Fig1}
\end{figure}

We consider a Bose Hubbard model on two-leg ladder  
as shown in Fig.\ref{Fig1}. We start considering the bosonic operator $b_{j_x,j_y}$ with  periodic/open boundary condition with
\begin{align*}
j_x & \equiv \frac 1 2 , \frac 3 2 \cdots, L- \frac 1 2 ,\ \text{mod }\, L \text{: even},\\
j_y& \equiv -\frac 1 2 ,\frac 1 2 ,\ \text{mod}\, 2.
\end{align*}
The Hamiltonian with periodic boundary condition is given by  (See Fig.\ref{Fig1})
\begin{eqnarray}
  H_{\rm BHM}&=&\sum^{\frac {L}{2}-1}_{j=0}(H_{j}+H_{j,\rm{int}}),
  \label{HBHM}
\end{eqnarray}
where 
\begin{eqnarray}
  H_j &=&H _j^\square + H_j^{\rm site},
\nonumber  \\
  H^\square_{j}&=&-J_v^1
  b^{\dagger}_{2j-\frac 1 2 ,-\frac 1 2 }b_{2j-\frac 1 2 ,+\frac 1 2 }
-J_v^2  b^{\dagger}_{2j+\frac 1 2 ,+\frac 1 2 }b_{2j+\frac 1 2 ,-\frac 1 2 }  
  \nonumber\\
  &&-J_h^1
  b^{\dagger}_{2j+\frac 1 2 ,-\frac 1 2 }b_{2j-\frac 1 2 ,-\frac 1 2 }
-J_h^2  b^{\dagger}_{2j-\frac 1 2 ,+\frac 1 2 }b_{2j+\frac 1 2 ,+\frac 1 2 })  
\nonumber\\
&&\qquad +\mbox{h.c.},
  \nonumber\\
  H_j^{\rm site} &=&
  \sum_{j_x=2j\pm \frac 1 2 }\bigg[\bigg(
    \sum_{j_y=\pm \frac 1 2 }
\frac{U}{2}n_{j_x,j_y}(n_{j_x,j_y}-1)
-\mu n_{j_x,j_y}\bigg)\nonumber \\
&&\qquad\qquad
+\frac{U_{in}}{2}n_{j_x,+\frac 1 2 }n_{j_x,-\frac 1 2 }\bigg],
  \nonumber\\
H_{j,{\rm int}} &=& -J_{\rm int}\sum_{j_y=\pm \frac 1 2 } b^{\dagger}_{2j-\frac 1 2 ,j_y}b_{2(j+1)-\frac 1 2 ,j_y}+\mbox{h.c.},  \nonumber
\end{eqnarray}
where $n_{j_x,j_y}=b^{\dagger}_{j_x,j_y}b_{j_x,j_y}$ and $L$ is the ladder length. $H_j$ is a Hamiltonian at the $j$-th plaquette and $J_{\rm int} $ connects them. $J_v^1,J_v^2,J_h^1,J_h^2\in\mathbb{R}$, are hopping amplitudes as shown in Fig.~\ref{Fig1}, $\mu$ is a chemical potential, and $U$-term is on-site interaction (between same internal states). The $U_{in}$-term represents interactions between the upper and lower chain. 
If the upper and lower chains are created by a different internal state atom in realistic experimental situations such as a synthetic ladder optical lattice \cite{Mancini2015,Stuhl2015}. The $U_{in}$-term can be regarded as interactions between different internal states and the $J_v^{1,2}$ -hopping is Rabi coupling.
We also comment that without the synthetic ladder the target system is also feasible in a real experimental system, 
such a bosonic plaquette optical lattice \cite{Paredes2008,Nascimbene2012} and also the model with synthetic gauge fields has been theoretically studied \cite{Greschner2016,Buser2020}.  
In what follows, we set $U=U_{in}$ and $\mu=U/2$ and a mean density ${\bar n}=\frac {1}{2L} \sum_{j_x,j_y} \langle n_{j_x,j_y} \rangle$  is used to specify the filling of the system. 
In most of our work, we focus on a strongly correlated regime, $|U|> |J^{1(2)}_{h}|, |J^{1(2)}_v|,|J_{\rm int}|$, practically $U = 20$ and consider soft-core bosons, where we expect that the ground state is always gapped unique and also any spontaneous symmetry breaking do not occur.

\section{Symmetry protection  and fractional quantization $Z_4$ Berry phase}
We introduce a $Z^{*}_2\times Z^{*}_2$ equivalence of the Berry phases associated with $Z_2\times Z_2$  symmetry (2 reflections) of
the ladder due to the hopping pattern and the form of the interaction.
This $Z_2\times Z_2$ symmetry can be employed to define a specific type of SPT phases. 
We then introduce a Berry phase by setting local twists on links. This $Z^{*}_2\times Z^{*}_2$ equivalence
leads to the $Z_4$ fractional quantization of the Berry phase. 

In what follows, we shall explain the $Z^{*}_2\times Z^{*}_2$ equivalence and introduce $Z_4$ Berry phase. Physical origin of the $Z_4$ quantization is special for the ladder compared with the previous studies
\cite{Hatsugai2006,Hirano2008,Maruyama2009,Hatsugai2011,Chepiga2013,Kariyado2015,Maruyama2018,Kariyado2018,Kawarabayashi2019,Mizoguchi2019,Araki2020,KY_2020,Otsuka2021,Bunney2022}.

\subsection{$Z_2\times Z_2$ symmetry}
The symmetry constraint to protect a SPT phase we discuss is a combination of two reflections with time-reversal. We consider two unitary operators, $U_x$ and $U_y$ for reflections shown in Fig.\ref{Fig1}
\begin{align*}
U_x H_{\rm BHM} U_x ^\dagger &=U_y H_{\rm BHM} U_y ^\dagger=H_{\rm BHM},
\end{align*}
where two unitary operators $U_x$ and $U_y$ operate for the boson operators as
\begin{align*}
U_x b_{j_x,j_y} U_x ^\dagger &= b_{-j_x,j_y},\quad U_x^2=1,\\
U_y b_{j_x,j_y} U _y ^\dagger &= b_{j_x,-j_y},\quad U_y^2=1.
\end{align*}
Note that this symmetry protection is respected for periodic boundary condition and also open boundary condition by cutting the link at the origin ($J_h=0$ at the links $(-\frac 1 2,\pm \frac 1 2  )-(+\frac 1 2,\pm \frac 1 2  )$).
It can guarantee the bulk-edge correspondence associated with $Z_4$ Berry phase as discussed later.

\begin{figure*}[t]
\includegraphics[width=14.5cm]{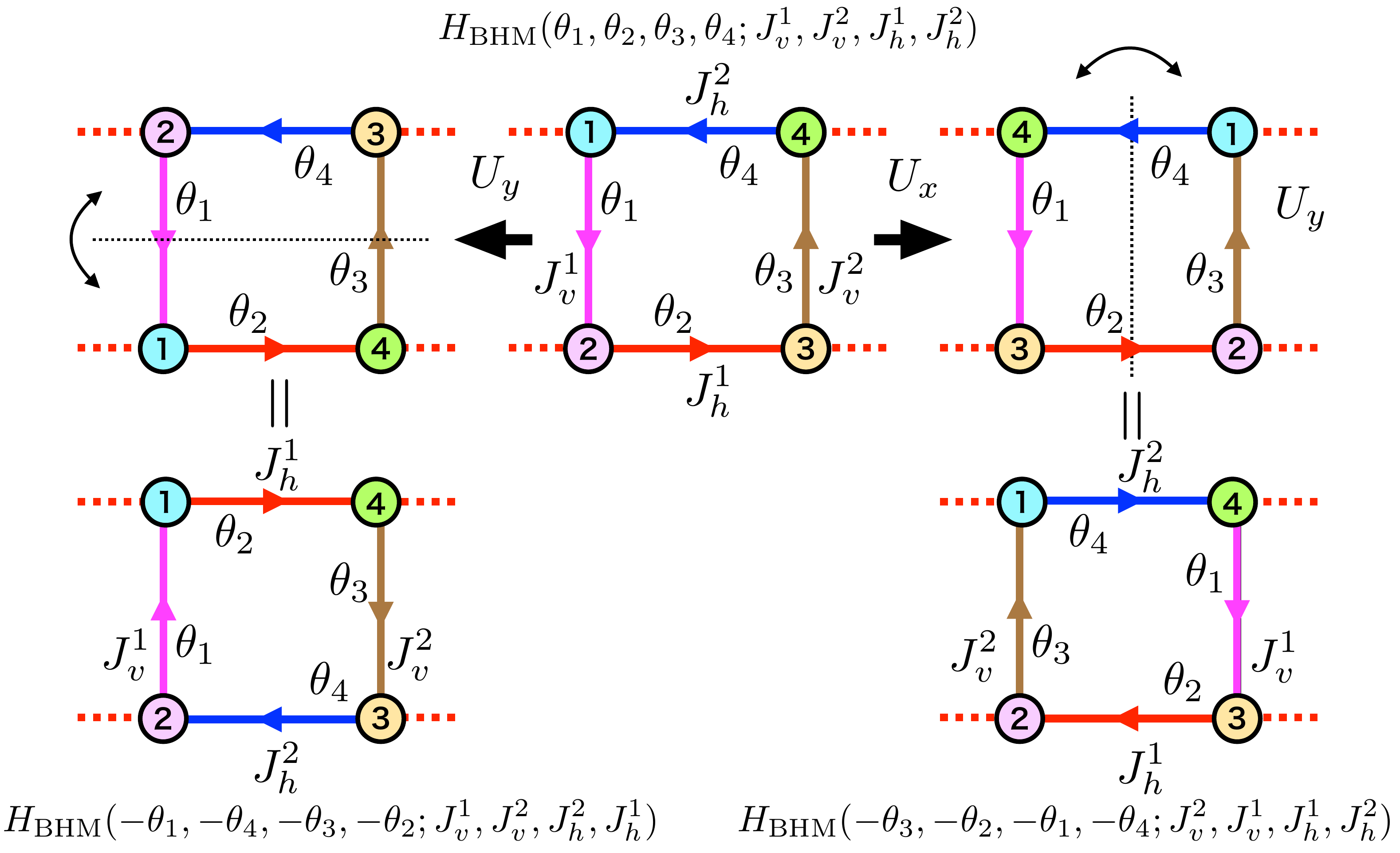}  
\caption{Two reflections for the Hamiltonian $H_{\rm BHM}$ after introducing the twists in a single plaquette at $j=0$.}
\label{Fig2}
\end{figure*}
\subsection{Quantized Berry phase as a topological local order parameter}
Let us introduce  quantized Berry phase
for a generic system by introducing a set of local twists as a parameter set. This type of the quantized Berry phase is a toplogical order parameter of the short-range entangled state \cite{Hatsugai2006,Hirano2008,Hatsugai2011,Kariyado2014,Kariyado2015,Araki2020}. 
The generic strategy is as follows. By introduction of the local twists as a set of parameters for the Hamiltonian, the Berry phase is defined for a many body ground state. 
Although, the Berry phase may take any value generically, 
one may impose symmetry constraints (with the twists), which induces quantization of the Berry phase. 
Due to the quantization, 
this quantized Berry phase can not be modified by small but finite perturbation. 
It implies that the quantized Berry phase is topologically stable and works local topological order parameter of the bulk. Since we need energy gap between the ground state (or set of ground states as a multiplet) and the other, this topological order parameter is only well defined for the gapped ground state. 
It picks up responses of the many body ground state against for the local twists as ``perturbation''. 
If the system is adiabatically modified (without gap closing) and decoupled into a set of local clusters, the system is short-range entangled. 
Since introduction of local gauge transformation of the bosonic/fermionic/spin operators inside some specific cluster does not affect the other clusters and induces twists only inside, it characterizes the locality. 
Using the twists, a Berry phase of the cluster can be defined assuming the ground state of the  cluster is unique. 
Apparently it is also considered as a Berry phase of the total system (although the cluster is decoupled). 
Also its value is easily evaluated since the twists are gauged out (See example below). 
Using the same twists, even with finite inter-cluster coupling, one may define a Berry phase associated with the many-body ground state. 
Note that the twists with the finite inter-cluster coupling can not be gauge out. In generically this Berry phase may take any value in modulo $2\pi$. 
However, due to symmetries the system possesses with the twists, the Berry phase may need to satisfy several constraints, 
which may result in the quantization. 
This is the quantized Berry phase and, then, 
it works as a topological order parameter of the system. 
Especially when the inter-cluster coupling is finite but weak enough, this quantized Berry phase is a topological order parameter of the short-range entangle state.

\subsection{$Z_4$ fractionalization in ladder}
Following this strategy, let us define $Z_4$ Berry phase
for the periodic boundary condition by introducing a set of $4$ twists,
$\Theta =(\theta_1,\theta_2,\theta_3,\theta_4)$, only in the plaquette at the origin ($j=0$) as shown in Fig.\ref{Fig1} and Fig.\ref{Fig2}.
The modified $\Theta $ dependent  Hamiltonian, $H_{\rm BHM}(\Theta )$, is given by replacing $H^\square_{0}\to H^\square_{0}(\Theta)$, 
\begin{align}
H^\square_{0}(\Theta )&=
-J_{v}^1 e^{-i \theta _1} b^{\dagger}_2 b_1
-J_{h}^1 e^{-i \theta _2} b^{\dagger}_3 b_2
\nonumber\\
&\qquad -J_{v}^2 e^{-i \theta _3} b^{\dagger}_4 b_3
-J_h^2 e^{-i \theta _4} b^{\dagger}_1 b_4
+\mbox{h.c.},
\label{plaquett-H}
\end{align}
where 
$b_1=b_{-\frac 1 2 ,+\frac 1 2 }$,
$b_2=b_{-\frac 1 2 ,-\frac 1 2 }$,
$b_3=b_{+\frac 1 2 ,-\frac 1 2 }$ and
$b_4=b_{+\frac 1 2 ,+\frac 1 2 }$.

\begin{figure}[t]
\begin{center} 
\includegraphics[width=7.5cm]{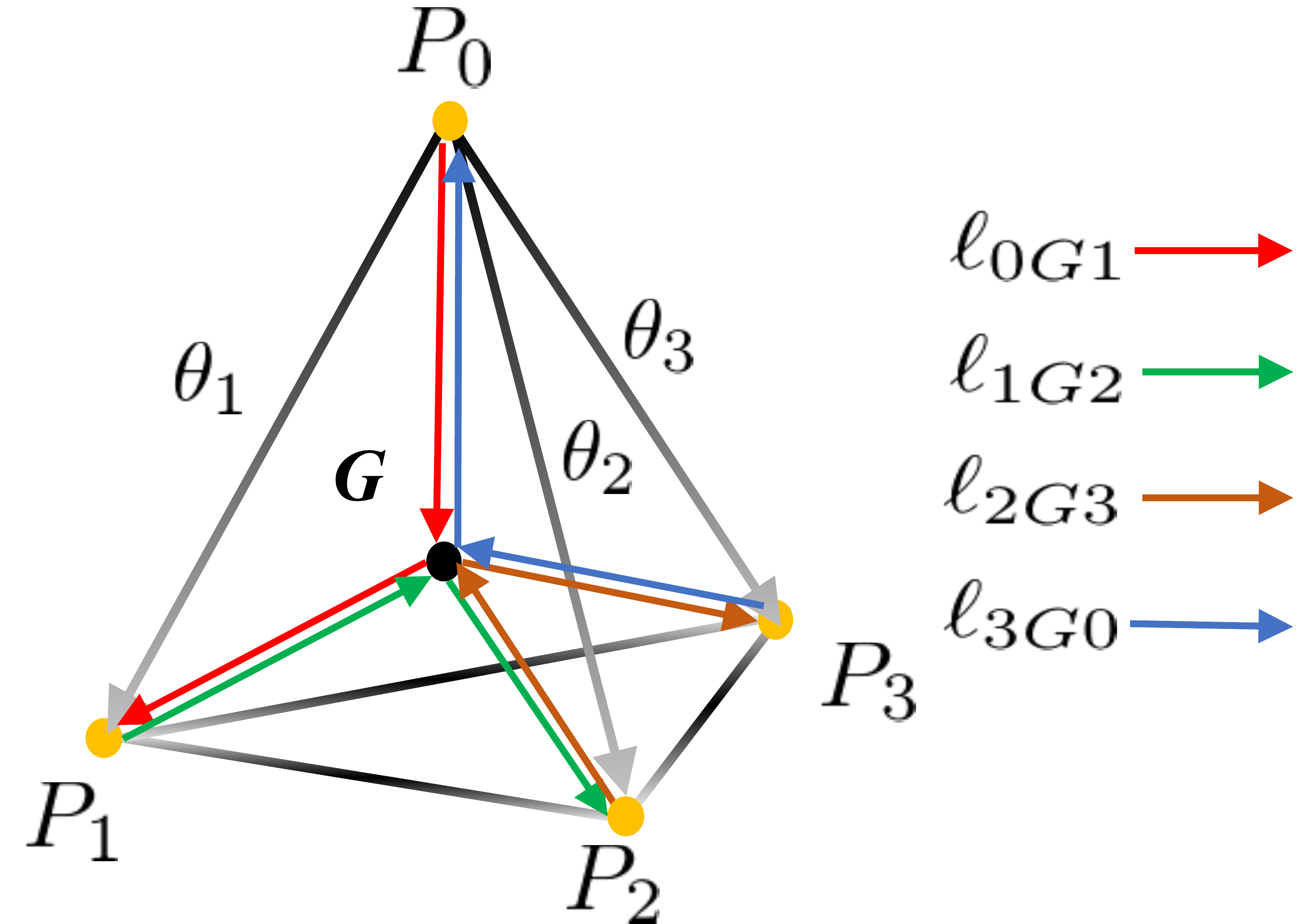}
\end{center} 
\caption{The three dimensional parameter space to define $Z_4$ Berry phases. By using an extended notation, the point in the space is specified by $(\theta_1,\theta_2,\theta_3,\theta_4 )$ where $\theta _1+\theta _2+\theta _3+\theta _4\equiv 0,\ (\text{mod }2\pi)$,
$P_0=(0,0,0,0)$, 
$P_1=(2\pi,0,0,-2\pi)$, 
$P_2=(0,2\pi,0,-2\pi)$, 
$P_3=(0,0,2\pi,-2\pi)$ 
and 
$G=(2\pi/4,2\pi/4,2\pi/4,2\pi/4)$. 
Identifying the equivalent points $P_0$, $P_1$, $P_2$ and $P_3$, this is a $3$-torus $T^3$.}
\label{Fig3}
\end{figure}

By imposing a constraint $\theta _1+\theta _2+\theta _3+\theta _4\equiv 0$, mod $2\pi$,
this set of the twists is specified by the point in the 3-torus,
$T^3=\{(\theta _1,\theta _2,\theta _3, \theta _4)| \theta _1+\theta _2+\theta _3+\theta _4\equiv 0,\ 
\theta _i\in \mathbb{R},\ \text{mod } 2\pi\}$ as shown in Fig.\ref{Fig3}.
We use this extended notation using 4 parameters to specify 3-torus, which is useful to discuss the $Z_4$ symmetry of the ladder (see below).

Identifying the equivalent points, any path connecting the vertices, $P_i$'s ($i=0,1,2,3$), define loops $\ell$ as shown in Fig.~\ref{Fig3}. Then assuming the ground state is unique (gap remains open) on the loop, the Berry phase is defined as
\begin{align*}
i \gamma _\ell &= \int_\ell \langle \psi| d\psi \rangle
\equiv \int_{s_i}^{s_f} ds\, \langle \psi(s)| \frac {d  }{d s } |\psi(s) \rangle,
\end{align*}
where $|\psi(\Theta) \rangle $ is a ground state of $H_{\rm BHM}(\Theta )$ ($  H_{\rm BHM}(\Theta )| \psi (\Theta )\rangle = | \psi(\Theta) \rangle E(\Theta )$
and  $s$ is any parameter to specify the loop, $\ell=\{\Theta(s)|\, s\in[s_i,s_f]\}$.
The $Z_4$ Berry phase
$\ell_\alpha $ is defined of 4 special paths as
\begin{eqnarray}
\ell_{0G1} &=& \vv{P_0G}+\vv {GP_1},\nonumber\\
\ell_{1G2} &=& \vv{P_1G}+\vv {GP_2},\nonumber\\
\ell_{2G3} &=& \vv{P_2G}+\vv {GP_3},\nonumber\\
\ell_{3G0} &=& \vv{P_3G}+\vv {GP_0},
\end{eqnarray}
where
\begin{eqnarray}
\vv{P_0G} &= \{(\theta ,\theta,\theta ,-3 \theta )|\ \theta \in(0,\frac {2\pi}{4}) \},\nonumber\\
\vv{P_1G} &= \{(-3\theta ,\theta,\theta , \theta )|\ \theta \in(0,\frac {2\pi}{4}) \},\nonumber\\
\vv{P_2G} &= \{(\theta ,-3\theta,\theta , \theta )|\ \theta \in(0,\frac {2\pi}{4}) \},\nonumber\\
\vv{P_3G} &= \{(\theta ,\theta,-3\theta , \theta )|\ \theta \in(0,\frac {2\pi}{4}) \}.
\end{eqnarray}

As for the twisted Hamiltonian, the reflections $U_x,U_y$, that make the untwisted Hamiltonian invariant,
operate as (See Fig.\ref{Fig2})

\begin{align*}
U_x & H_{\rm BHM}(\theta _1,\theta _2,\theta _3,\theta _4;J_v^1,J_v^2,J_h^1,J_h^2) U_x ^{-1}\\
&=  H_{\rm BHM}(-\theta _3,-\theta _2,-\theta _1,-\theta _4;J_v^2,J_v^1,J_h^1,J_h^2),\\
U_y & H_{\rm BHM}(\theta _1,\theta _2,\theta _3,\theta _4;J_v^1,J_v^2,J_h^1,J_h^2) U_y ^{-1}\\
&= H_{\rm BHM}(-\theta _1,-\theta _4,-\theta _3,-\theta _2;J_v^1,J_v^2,J_h^2,J_h^1),
\end{align*} 
where relevant parameter dependence is explicitly shown.
If $J_v^1=J_v^2$, the twisted Hamiltonian on the loop is mapped as,
\begin{eqnarray}
\Xi _x \left.H_{\rm BHM}(\Theta )\right|_{\vv {P_0G}} \Xi _x ^{-1} &=& \left.H_{\rm BHM}\right|_{\vv {P_0G}}(\Theta ),\nonumber \\
\Xi _x \left.H_{\rm BHM}(\Theta )\right|_{\vv {P_1G}} \Xi _x ^{-1} &=& \left.H_{\rm BHM}\right|_{\vv {P_3G}}(\Theta ),\nonumber \\
\Xi _x \left.H_{\rm BHM}(\Theta )\right|_{\vv {P_2G}} \Xi _x ^{-1} &=& \left.H_{\rm BHM}\right|_{\vv {P_2G}}(\Theta ),\nonumber \\
\Xi _x \left.H_{\rm BHM}(\Theta )\right|_{\vv {P_3G}} \Xi _x ^{-1} &=& \left.H_{\rm BHM}\right|_{\vv {P_1G}}(\Theta ),
\end{eqnarray}
where
\begin{align*}
\Xi _x &=  U_x K,\ \Xi _y =  U_y K.
\end{align*}
It implies
\begin{align*} 
\Xi _x \left.H_{\rm BHM}(\Theta )\right|_{\ell_{0G1}} \Xi _x ^{-1} &= \left.H_{\rm BHM}\right|_{\ell_{0G3}}(\Theta ),\\
\Xi _x \left.H_{\rm BHM}(\Theta )\right|_{\ell_{1G2}} \Xi _x ^{-1} &= \left.H_{\rm BHM}\right|_{\ell_{3G2}}(\Theta ),\\
\Xi _x \left.H_{\rm BHM}(\Theta )\right|_{\ell_{2G3}} \Xi _x ^{-1} &= \left.H_{\rm BHM}\right|_{\ell_{2G1}}(\Theta ),\\
\Xi _x \left.H_{\rm BHM}(\Theta)\right|_{\ell_{3G0}} \Xi _x ^{-1} &= \left.H_{\rm BHM}\right|_{\ell_{1G0}}(\Theta ).
\end{align*}
Note that, in general, 
as for the parameter independent anti-unitary operator
$\Xi =U K$, where $U U ^\dagger =1$,
the Berry connection of the state $|\psi^\Xi \rangle =\Xi|\psi \rangle $ is
$\langle \psi^\Xi | d\psi^\Xi \rangle= -  \langle \psi | d\psi \rangle $
since $\langle \psi | d\psi \rangle $ is pure imaginary.
Then using abbreviation,
$  \gamma _0 \equiv \gamma _{\ell_{0G1}}$,
$  \gamma _1 \equiv \gamma _{\ell_{1G2}}$,
$  \gamma _2 \equiv \gamma _{\ell_{2G3}}$, 
$  \gamma _3 \equiv \gamma _{\ell_{3G0}}$,
it is further written as
\begin{align*}
-\gamma _0 &\equiv \gamma _0+\gamma _1+\gamma _2,\\
-  \gamma _1 &\equiv -\gamma _2,\\
-  \gamma _2 &\equiv -\gamma _1,\\
-  \gamma _3 &\equiv -\gamma _0.
\end{align*}
By using the apparent relation, 
$\gamma _0+\gamma _1+\gamma _2+\gamma _3\equiv 0$ mod $2\pi$, due to the cancellation of the $4$ paths,
it is summarized as
\begin{align}
\gamma _0 &\equiv \gamma _3,\label{g0g3BP}\\
\gamma _1 &\equiv \gamma _2.
\label{g1g2BP}
\end{align}
It naturally implies a partial quantization,
\begin{align}
\gamma _i+ \gamma _j &=0,\ \pi\ (\text{mod}\, 2\pi),
\label{q-case1}
\end{align}
for any $ i\ne j$ except $(i,j)= (0,3), (1,2)$.

Similarly if $J_h^1=J_h^2$, we have following relations,
\begin{eqnarray}
\Xi _y \left.H_{\rm BHM}(\Theta )\right|_{\vv {P_0G}} \Xi _y ^{-1} &=& \left.H_{\rm BHM}\right|_{\vv {P_2G}}(\Theta ),\nonumber \\
\Xi _y \left.H_{\rm BHM}(\Theta )\right|_{\vv {P_1G}} \Xi _y ^{-1} &=& \left.H_{\rm BHM}\right|_{\vv {P_1G}}(\Theta ),\nonumber \\
\Xi _y \left.H_{\rm BHM}(\Theta )\right|_{\vv {P_2G}} \Xi _y ^{-1} &=& \left.H_{\rm BHM}\right|_{\vv {P_0G}}(\Theta ),\nonumber \\
\Xi _y \left.H_{\rm BHM}(\Theta )\right|_{\vv {P_3G}} \Xi _y ^{-1} &=& \left.H_{\rm BHM}\right|_{\vv {P_3G}}(\Theta ).
\end{eqnarray}
It implies
\begin{align*} 
\Xi _y \left.H_{\rm BHM}(\Theta )\right|_{\ell_{0G1}} \Xi _y ^{-1} &= \left.H_{\rm BHM}\right|_{\ell_{2G1}}(\Theta ),\\
\Xi _y \left.H_{\rm BHM}(\Theta)\right|_{\ell_{1G2}} \Xi _y ^{-1} &= \left.H_{\rm BHM}\right|_{\ell_{1G0}}(\Theta ),\\
\Xi _y \left.H_{\rm BHM}(\Theta)\right|_{\ell_{2G3}} \Xi _y ^{-1} &= \left.H_{\rm BHM}\right|_{\ell_{0G3}}(\Theta ),\\
\Xi _y \left.H_{\rm BHM}(\Theta)\right|_{\ell_{3G0}} \Xi _y ^{-1} &= \left.H_{\rm BHM}\right|_{\ell_{3G2}}(\Theta ).
\end{align*}
and then
\begin{align*}
- \gamma _0 &\equiv -\gamma _1\\
- \gamma _1 &\equiv -\gamma _0,\\
- \gamma _2 &\equiv \gamma _0+\gamma _1+\gamma_2\equiv - \gamma _3,\\
-  \gamma _3 &\equiv -\gamma _2.
\end{align*} 
It is summarized as
\begin{align}
\gamma _0 &\equiv \gamma _1,  \\
\gamma _2 &\equiv \gamma _3.
\end{align}
It naturally implies a partial quantization,
\begin{align}
\gamma _i+ \gamma _j &=0,\ \pi\ (\text{mod}\, 2 \pi), 
\label{q-case2}
\end{align}
for any $ i\ne j$ except $(i,j)= (0,1), (2,3)$.

If the ladder satisfies the full $Z_2\times Z_2$ symmetries,
$J_v^1=J_v^2$ and $J_h^1=J_h^2$, supplemented with the  constraint
$\gamma _0+\gamma _1+\gamma _2+\gamma _3=0$,
\begin{eqnarray}
\gamma_0\equiv \gamma_1 \equiv \gamma_2 \equiv  \gamma_3= 2\pi \frac{n}{4},(n \in \mathbb{Z})\ (\text{mod}\, 2 \pi).
\label{Z4_BP}
\end{eqnarray}
This is the $Z_4 $ quantization of the ladder. We call it $Z_4$ Berry phase \cite{Hatsugai2011,HK2023}.

\subsection{Plaquette limit}
The quantized values, Eq.(\ref{q-case1}) , Eq.(\ref{q-case2}) and  Eq.(\ref{Z4_BP}) are fixed explicitly
if the system is adiabatically connected to a set of  decoupled plaquette ($J_{\rm int}=0$).

As for the $Z_4$ Berry phase,
it is enough to consider a single plaquette in this case.
The twists $\Theta $ are 
gauge out, (or they are induced) by the gauge transformation
(See Eq.(\ref{plaquett-H}))
\begin{eqnarray}
H_{\rm BHM}(\Theta) &=& U_\Theta H_{\rm BHM} U_\Theta ^{-1},\quad \text{if } J_{\rm int}=0,\nonumber\\
U_{\Theta } &=& e^{-i\phi_1 n_{1}} e^{-i\phi_2 n_{2}}e^{-i\phi_3 n_{3}}e^{-i\phi_4 n_{4}}, \nonumber
\end{eqnarray}
where note that $ U_{\Theta }b_{j }U^\dagger_{\Theta } = e^{+i \phi_j }b_{j}$, ($j=1,2,3,4$).
$\phi_1(\Theta)=0$,
$\phi_2(\Theta )=\theta_1$,
$\phi_3 (\Theta )=\theta_1+\theta_2$,
$\phi_4 (\Theta )=\theta_1+\theta_2+\theta_3$.
It implies that $|\psi(\Theta )\rangle=U_\Theta |\psi(0)\rangle$.

Noting that $[n_i,U_\Theta ]=0$, the Berry connection and the Berry phase
are given as
\begin{align*}
\langle \psi| d\psi \rangle &= \langle \psi(0)| U_\Theta  ^\dagger d U_\Theta |\psi(0) \rangle
=- i \sum_{i=1}^4 d \phi_i \langle n_i \rangle_0,\\
\gamma _\ell &=   -i\int_\ell\langle \psi| d\psi \rangle
=\sum_i \langle n_i \rangle_0 \Delta \phi_i,
\end{align*}
where $\langle n_i \rangle_0 = \langle \psi(0)|n_i|\psi(0) \rangle  $ and
$\Delta \phi_i=\phi_i(\Theta (s))\big|_{s_i}^{s_f}$.

As for the canonical loops, it is
\begin{align*}
\gamma _0 &= -2\pi ( \langle n_2 \rangle _0+ \langle n_3 \rangle _0
+ \langle n_4 \rangle _0),\\
\gamma _1 &= 2\pi  \langle n_2 \rangle _0,\\
\gamma _2 &= 2\pi  \langle n_3 \rangle _0,\\
\gamma _3 &= 2\pi  \langle n_4 \rangle _0.
\end{align*}
We assume that
the ground state of the total system (and thus, that of the plaquette as well) is unique. Then total number of boson in the plaquette is positive integer
$M=\sum_{i=1}^4 \langle n_i \rangle _0$.

When $J_v^1=J_v^2$, due to the $Z_2$ invariance by $U_x$,
$\langle n_1 \rangle _0=\langle n_4 \rangle _0$
and
$\langle n_2 \rangle _0=\langle n_3 \rangle _0$.  It implies
\begin{align*}
\gamma _0 &=-2\pi(M- \langle n_1 \rangle _0)\equiv  2\pi  \langle n_1 \rangle _0  \\
& =  2\pi  \langle n_4 \rangle _0=
\gamma _3,\ \text{mod }2\pi,\\
\gamma _1 &= \gamma _2.
\end{align*}
Also due to this $Z_2$,
$\langle n_1 \rangle _0+\langle n_2 \rangle _0=
\langle n_3 \rangle _0+\langle n_4 \rangle _0=
M/2 $, that implies the partial quantization,
\begin{align*}
\gamma _0 + \gamma _1
&\equiv
\gamma _2 + \gamma _3  \equiv \frac {M}{2},\ (\text{mod } 2\pi).
\end{align*}

Similarly, when $J_h^1=J_h^2$, due to the $Z_2$ invariance by $U_y$,
$\langle n_1 \rangle _0=\langle n_2 \rangle _0$
and
$\langle n_3 \rangle _0=\langle n_4 \rangle _0$.  
It implies
\begin{align*}
\gamma _0 &=-2\pi(M- \langle n_1 \rangle _0)\equiv  2\pi  \langle n_1 \rangle _0 \\
& =  2\pi  \langle n_2 \rangle _0=\gamma _1,\ \text{mod }2\pi,\\
\gamma _2 &= \gamma _3.
\end{align*}
Also $\langle n_1 \rangle _0+\langle n_4 \rangle _0= \langle n_2 \rangle _0+\langle n_3 \rangle _0= M/2 $, that implies the partial quantization,
\begin{align*}
\gamma _0 + \gamma _3
&\equiv
\gamma _1 + \gamma _2  \equiv \frac {M}{2},\  (\text{mod } 2\pi).
\end{align*}

Then when the system is $Z_2\times Z_2$ invariant, $J_v^1=J_v^2$ and $J_h^1=J_h^2$, we have $Z_4$ quantization as
\begin{align*}
\gamma _0 & \equiv   \gamma _1 \equiv   \gamma _2 \equiv   \gamma _3 \equiv \frac {M}{4},\ (\text{mod } 2\pi).
\end{align*}
These quantized Berry phases are adiabatic invariants.

\subsection{Another $Z_2$ Berry phase}
We also consider a set of  twists in $H_0$
by assuming the $Z_2$ symmetry due to $U_x$, that is, $J_v^1=J_v^2=J_v$ as
\begin{align} 
H^\square_{0}(\theta )&=
-J_{v} b^\dagger_2 b_1
-J_{h}^1 e^{-i \theta } b^\dagger_3 b_2
\nonumber\\
&\qquad -J_{v}  b^\dagger_4 b_3
-J_h^2 e^{-i \theta } b^\dagger_1 b_4
+\mbox{h.c.}.
\end{align} 
We also consider a Berry phase $\gamma ^{Z_2}$ associated with this set of twists.
Due to the $Z_2$ symmetry, it is quantized into $Z_2$ as 
\begin{align*}
  \gamma ^{Z_2} &\equiv 0,\pi,\  (\text{mod }2\pi).
\end{align*}
This is due to the symmetry constraint
\begin{align*}
  \gamma ^{Z_2} &\equiv  - \gamma ^{Z_2} .
\end{align*}

A dimer limit ($J_v^1=J_v^2=0$, $J_{\rm int}=0$) is the
decoupled limit for this case.
\begin{eqnarray}
H_{\rm BHM}(\theta)&=& U_\theta H_{\rm BHM} U_\theta ^{-1},\quad \text{if}\: J_v=J_{\rm int}=0,\\
U_{\theta } &=& e^{-i \theta n_{1}} e^{-i \theta  n_{3}}.
\end{eqnarray}
In this decoupled case,
\begin{align*}
\langle \psi| d\psi \rangle &= \langle \psi(0)| U_\theta  ^\dagger d U_\theta |\psi(0) \rangle = -i d \theta ( \langle n_1 \rangle_0+\langle n_3 \rangle_0),\\
\gamma ^{Z_2}&= -i\int_\ell \langle \psi| d\psi \rangle=2\pi( \langle n_1 \rangle_0+\langle n_3 \rangle_0).
\end{align*}
Further
$\langle n_1 \rangle_0=\langle n_4 \rangle_0$ and
$\langle n_2 \rangle_0=\langle n_3 \rangle_0$
due to the $Z_2$ invariance by $U_x$.
It implies
$\langle n_1 \rangle_0+\langle n_3 \rangle_0=
\langle n_2 \rangle_4+\langle n_3 \rangle_0=M/2
$.
Then adiabatic continuation
to this dimer limit guarantees
\begin{align*}
\gamma ^{Z_2} &\equiv 2\pi \frac {M}{2}\equiv \pi M,\ (\text{mod } 2\pi).
\end{align*}

\section{Numerical evaluation of the $Z_4$ SPT phase}

In the previous section, we showed the $Z_4$ fractional quantization of the $Z_4$ Berry phase by the $Z_2\times Z_2$ symmetry. 
We herein turn to the numerical demonstration of its fractional quantization in the Hamiltonian $H_{\rm BHM}$ by using diagonalization \cite{Quspin} for various parameter conditions. 
The fractional quantization of the the $Z_4$ Berry phase is the signal of the presence of the bulk SPT phase protected by the $Z_2\times Z_2$ symmetry. 

In what follows, we set $J_v^1=J_v^2=J_v$ and $J_h^1=J_h^2=J_h$. 
We introduce a dimerization parameter $\delta J$ as $J_{\rm int}=1-\delta J$, $J_h=\delta J$ and set $J_v=1$. 
This setting preserves the $Z_2\times Z_2$ symmetry. 
We focus on $U=20$ and consider soft-core bosons.

\begin{figure}[t]
\begin{center} 
\includegraphics[width=8.5cm]{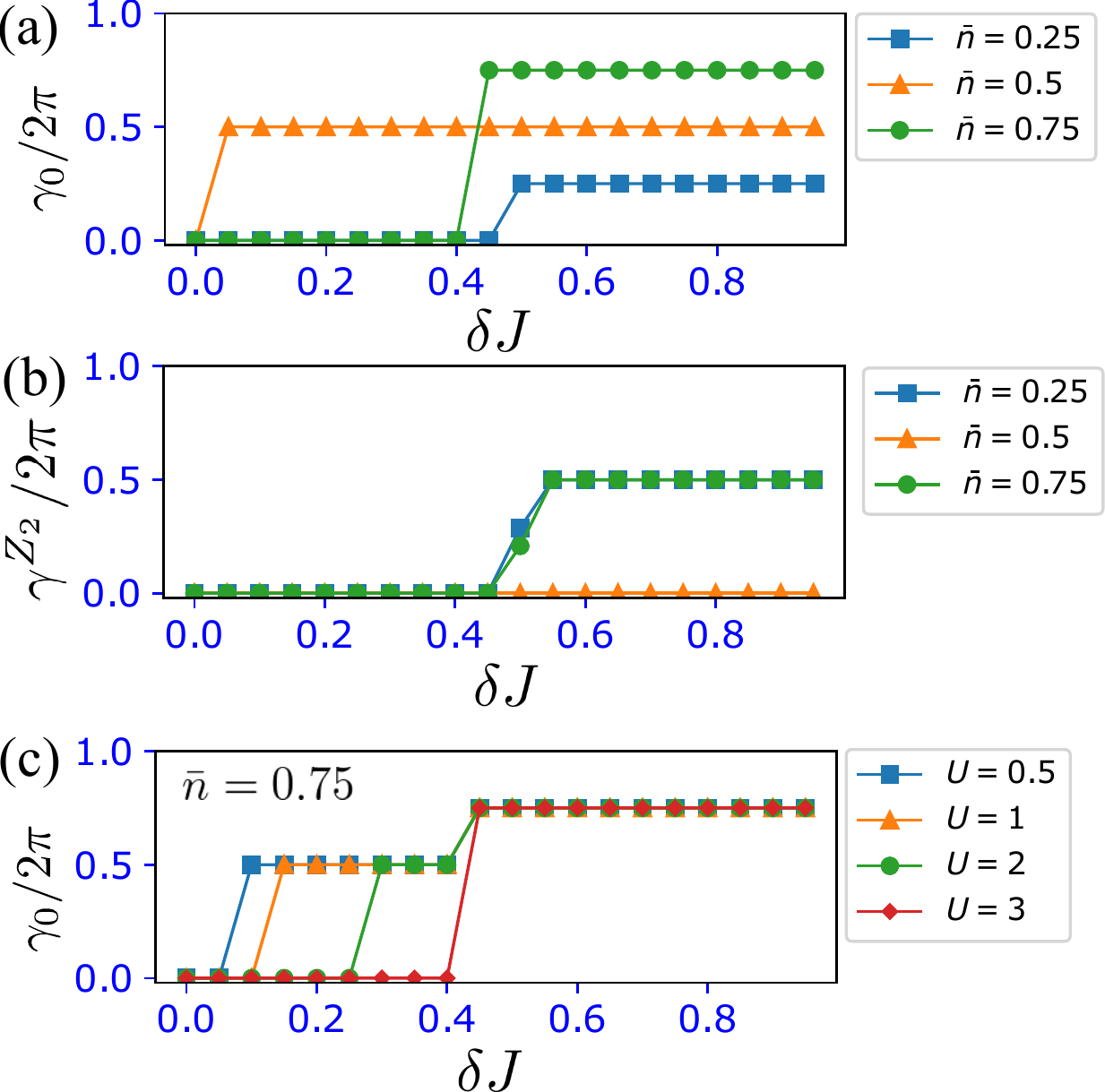}  
\end{center} 
\caption{
(a) Behavior of the $Z_4$ Berry phase for various mean densities with $U=20$. 
(b) Behavior of the $Z_2$ Berry phase for various mean densities with $U=20$. 
(c) Behavior of the $Z_4$ Berry phase for various $U$ with ${\bar n}=0.75$.
In the numerical diagonalization, the maximum occupation number of the boson on a site is truncated up to four. 
For $U\ll 1$, a superfluid phase appears where the gap is very small and the value of the Berry phase is unstable.
}
\label{Fig4}
\end{figure}

As varying $\delta J$ for ${\bar n}=0.25$, $0.5$ and $0.75$, the $Z_4$ Berry phase behaves as shown in Fig.~\ref{Fig4} (a). 
For $\delta J > 0.5$ and ${\bar n}=0.25$ and $0.75$, we observe the fractional quantization $\gamma_0/2\pi=1/4$ and $3/4$, 
and the $Z_4$ Berry phase captures clear topological phase transitions at $\delta J=0.5$. 
On the other hand, for ${\bar n}=0.5$, the finite fractional quantization $\gamma_0/2\pi =2/4$ appears even for a finite $\delta J$. The reason is that two dimer states resided on $J_v$-links on a plaquette is adiabatically connected to a plaquette cluster state. 
These results of the fractional quantization are the signal of the presence of the bulk SPT phases of the $Z_2\times Z_2$ symmetry. 
Also note that the quantization value of the $Z_4$ Berry phase is related to the mean density $\gamma_0/2\pi={\bar n}$, which is expected by considering a decoupled plaquette limit.

We then calculate the $Z_2$ Berry phase. 
Since the model of $H_{\rm BHM}$ has inversion symmetry, 
thus, the $Z_2$ Berry phase can capture a topological phase transition from $\gamma^{Z_2}=0$ to $\gamma^{Z_2}=\pi$ \cite{Kariyado2015}. Figure \ref{Fig4} (b) is the behavior of the $Z_2$ Berry phase.  
We observe that for ${\bar n}=0.25$ and $0.75$ cases, the $Z_2$ Berry phase characterizes a topological phase transition and its topological phase but for ${\bar n}=0.5$, the $Z_2$ Berry phase does not capture the presence of the bulk SPT phase.
 
We observe effects of the interaction $U$. 
The $Z_{4}$ Berry phase exhibits an interesting behavior for ${\bar n}=0.75$ as shown in Fig.~\ref{Fig4} (c). 
As varying $\delta J$, we observe an intermediate plateau phase with $\gamma_0/2\pi=1/2$ for a moderate $U$ and $\delta J$, this behavior is a specific character in bosonic system \cite{Deng2014,KSI2017}. This intermediate phase we find is interesting in that, this state is not adiabatically connected to the state of the plaquette limit, which exhibits $\gamma_0/2\pi=3/4$ state.

\begin{figure}[t]
\begin{center} 
\includegraphics[width=8cm]{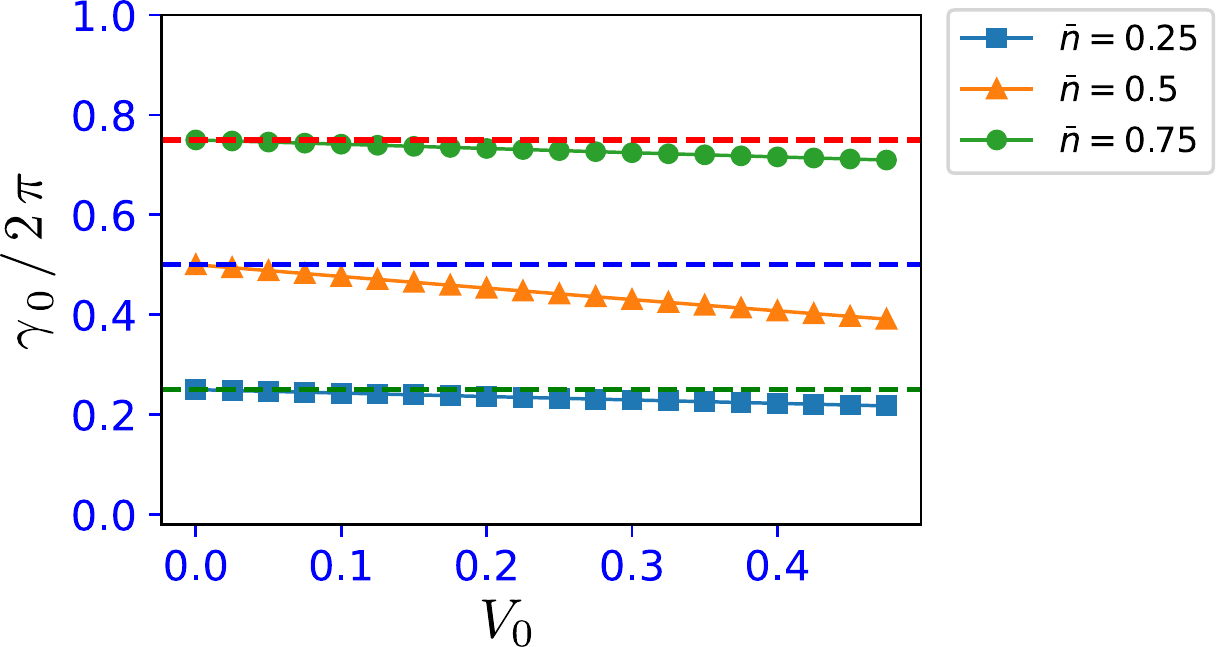}  
\end{center} 
\caption{
Effect of the perturbation breaking the $Z_2\times Z_2$ symmetry. 
We set $\delta J=0.8$ and $U=20$.
The red, blue and green dashed lines represent the ideal quantization values $\gamma_0/2\pi=1/4$, $2/4$ and $3/4$.}
\label{Fig5}
\end{figure}

Next, to observe the importance of the $Z_2\times Z_2$ symmetry for the presence of SPT phase, we here introduce a symmetry breaking potential. 
As a perturbation, we add a potential,
$$
V_p=V_0\sum^{\frac {L}{2}-1}_{j=0}(n_{2j-\frac 1 2 ,-\frac 1 2 }+n_{2j+\frac 1 2 ,+\frac 1 2 }),
$$ 
which breaks the $Z_2\times Z_2$ symmetry. 
We expect that as increasing $V_0$, 
the fractional quantization of the $Z_4$ Berry phase is collapsed. 
In fact, we verify this expectation as shown in Fig.~\ref{Fig5}. For each densities ${\bar n}$, the quantization continuously breaks down as increasing $V_0$. 
This result indicates the $Z_2\times Z_2$ symmetry is crucial for existing the SPT phase in the bosonic system.

An interesting behavior of the $Z_4$ Berry phase in the system with {\it only} $U_x$  symmetry is observed ($U_y$ is broken).
From Eqs.~(\ref{g0g3BP}) and (\ref{g1g2BP}), a certain relation of $Z_4$ Berry phase exists,  
\begin{eqnarray}
\gamma_0+\gamma_1&=&\gamma_2+\gamma_3.
\end{eqnarray}
Combined with the relation $\sum_{\alpha}\gamma_{\alpha}=0$ $\pmod {2\pi}$, one obtain
\begin{eqnarray}
\gamma_0+\gamma_1&=&0,\:\:\pi \pmod {2\pi},\nonumber \\
\gamma_2+\gamma_3&=&0,\:\:\pi \pmod {2\pi}.
\label{BP_con3}
\end{eqnarray}
The sum of the two $Z_4$ Berry phases is quantized.
We verify this unconventional quantization.
To this end, we modify the parameters 
$J_{\rm int}$ and $J_h$ as 
\begin{eqnarray}
&&J_{\rm int} \to \left\{
\begin{array}{ll}
 1-\delta J\equiv J^{2}_{\rm int}, &(\mbox{upper chain})\\
 1-(\delta J-0.1)\equiv J^{1}_{\rm int}, &(\mbox{lower chain})\\
\end{array}
\right.
\end{eqnarray}
\begin{eqnarray}
J^2_h=\delta J,\:\: J^1_h&=&\delta J-0.1.
\end{eqnarray}

\begin{figure}[t]
\begin{center} 
\includegraphics[width=7cm]{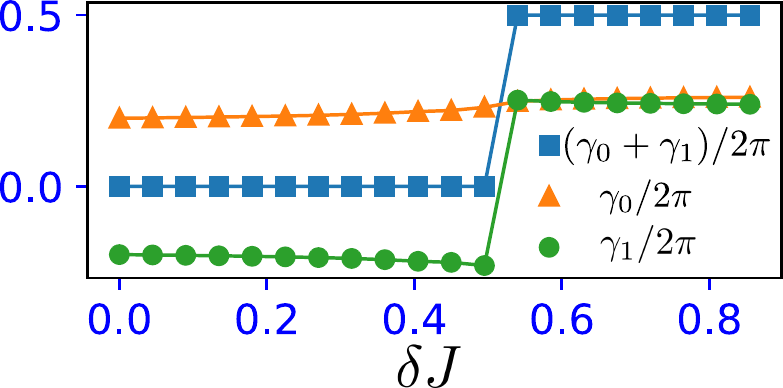}  
\end{center} 
\caption{$\delta J$-dependence of $Z_4$ Berry phase in the system with 
only $U_x$ symmetry. Even for $\delta J> 0.4$, $\gamma_0/2\pi$, $\gamma_1/2\pi \neq 1/4$. 
}
\label{Fig6}
\end{figure}

The Hamiltonian $H_{\rm BHM}$ is no longer invariant for the $Z_2$ symmetry $U_y$ and invariant only for the $Z_2$ symmetry $U_x$. 
For ${\bar n}=0.25$, we demonstrate the quantization of the sum of the $Z_4$ Berry phases. 
Figure~\ref{Fig6} is the numerical result. 
Depending on the value of $\delta J$, 
the sum $\gamma_0+\gamma_1$ takes $0$ or $\pi$ and 
we observe a clear phase transition while each $\gamma_0$ and $\gamma_1$ take some fractional values for any $\delta J$ or do not take $2\pi/4$. 
Hence, we confirm the $Z_2$ quantization of the sum of the $Z_4$ Berry phase as shown in Eq.~(\ref{BP_con3}). 

In addition, we expect that the bulk SPT phases appear in spinless free fermion system. See Appendix, we confirmed the presence of the SPT phase, which are also characterized by the fractional quantization of the $Z_4$ Berry phase. 

\begin{figure}[t]
\begin{center} 
\includegraphics[width=7cm]{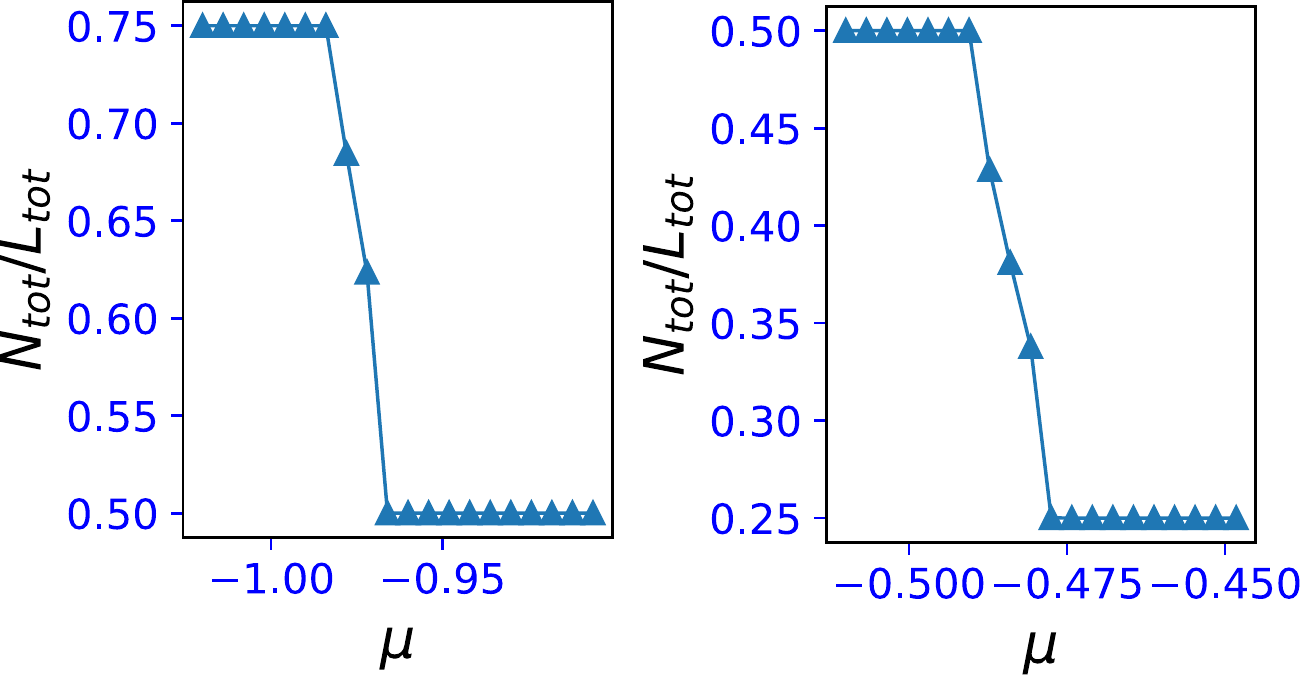}  
\end{center} 
\caption{$\mu$-dependence of total density in the system with periodic boundary condition.
The total number of the lattice site is $L_{tot}=64$ ($L=32$) and $N_{tot}$ is the total particle number. We set $J_{\rm int}=0.1$, $J_{h}=0.9$ and $U=20$.}
\label{Fig7}
\end{figure}

Before going to the next section, we show the DMRG calculation allowing the change of the particle number in the system to consolidate the presence of the bulk SPT phases as shown in Fig.\ref{Fig5} (a). We employ the DMRG algorithm by using TeNPy library \cite{TeNPy}. In the calculation, we vary the chemical potential $\mu$ in the system with the periodic boundary condition. The results are shown in Fig.~\ref{Fig7}, where the same parameter condition is set as in Fig.\ref{Fig4} (a).
We find some plateaus with the total particle number constant. 
The results indicate that the density on each plateau corresponds to that of each bulk SPT phases as shown in Fig.\ref{Fig7} (a) and \ref{Fig7} (b), each state on each plateau is gapped and incompressible. 
This is reminiscent of the appearance of the magnetization plateaus \cite{Oshikawa1997,Kariyado2015} and the density plateaus of Mott insulator in the conventional Bose Hubbard model \cite{Fisher1989}.

\section{Edge states in bosonic SPT phases}

In general, the bulk-edge correspondence is to read out the information of the bulk from various edge states under various forms of edge. Conversely, if a non-trivial bulk state exists, then one introduces an edge in the system and some edge state appears. It is possible to predict the form of edge state from inferring what states appear by cutting the decoupled plaquette in this system. 

For the bosonic system in this work, if we believe the presence of the bulk-edge correspondence, some bosonic edge state can appear under some edge shape. However, it is difficult to directly deduce the presence and detailed properties of edge state since our considering bosonic system is complex due to the ladder geometry, the presence of the interactions, the vertical hoppings and soft-core boson nature. It is difficult to identify some symmetries that would characterize the appearance of edge states as they appear in free fermions (cf. chiral symmetry). 
Even in this situation, when we set an experimentally feasible edge shape of interest, we can confirm numerically on a case-by-case what edge states appear, at least. Thus, we perform such a study in the following.

To detect some edge state, we also employ DMRG algorithm \cite{TeNPy}. 
We here impose an open boundary condition and analyze the two cases: 
(i) a vertical edge case and (ii) a diagonal edge case. 
For both of cases, we employ the simulation allowing the change of particle number in the system (grand-canonical), that is, vary the chemical potential $\mu$ and observe density properties of both bulk and edges. 
We expect that edge states (if appear) are different from that of a conventional free-fermion system with a certain topological phase. 
In particular, we comment that it is care that in general, bosonic systems with open boundary do not necessarily have zero-energy edge state, e.g. due to the lack of chiral symmetry. Some interacting bosonic systems have already reported it and numerically verified in the context of the topological Mott insulator \cite{Grusdt2013} and Haldane insulator \cite{Stumper2020}.

\subsection{Vertical edge case}
\begin{figure}[t]
\begin{center} 
\includegraphics[width=8cm]{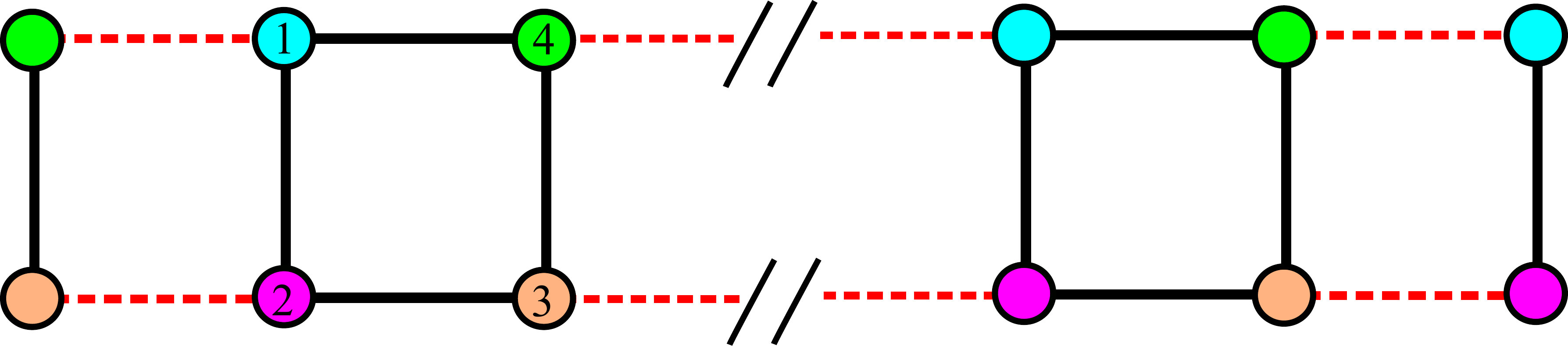}  
\end{center} 
\caption{Schematic image of the introduction of a vertical edge. 
This open boundary system is invariant under both $U_x$ and $U_y$.}
\label{Fig8}
\end{figure}

We first focus on the system with vertical edge as shown in Fig.~\ref{Fig8}. 
The edge is simply introduced by cutting a single plaquette at $j=0$ of the periodic system. The boundary preserves both $U_x$ and $U_y$ symmetries.

\begin{figure}[t]
\begin{center} 
\includegraphics[width=7cm]{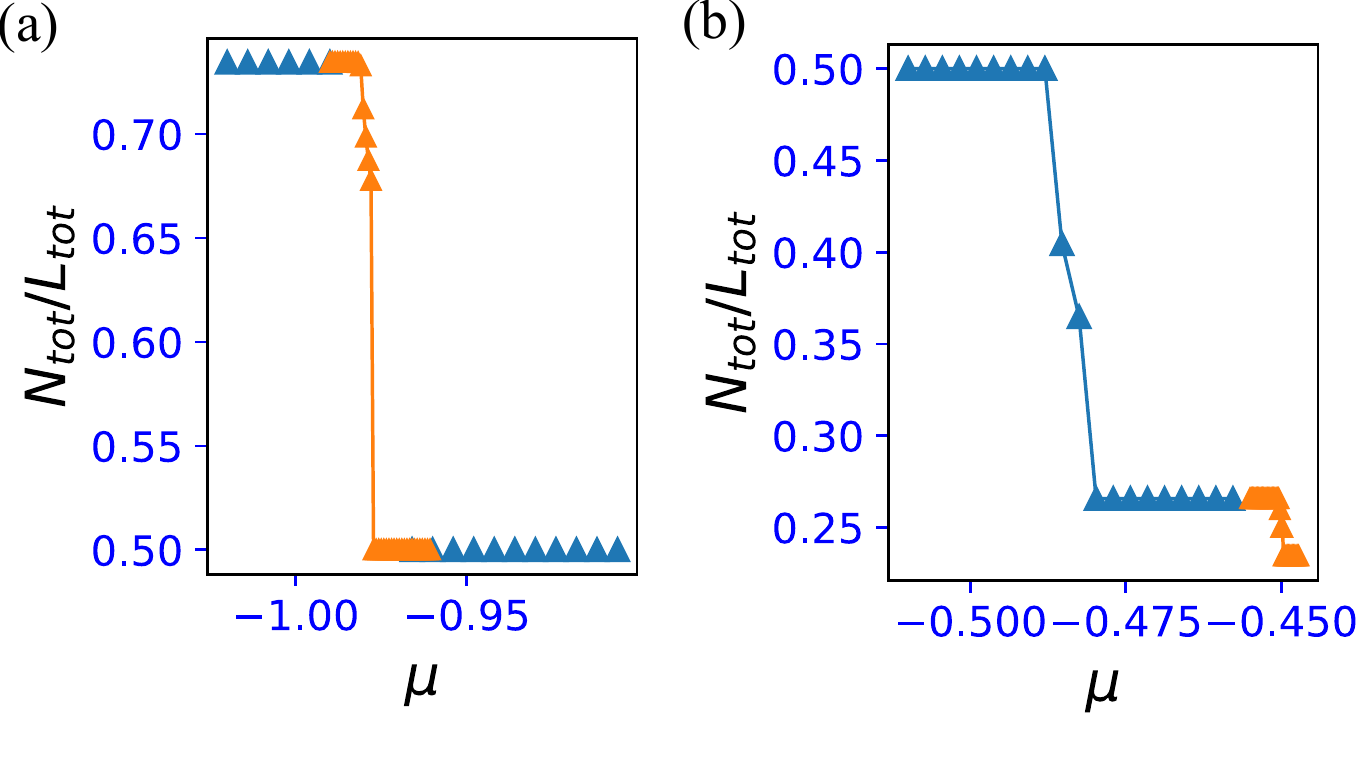}  
\end{center} 
\caption{$\mu$-dependence of total density, $N_{tot}/L_{tot}$.
$L_{tot}=64$ ($L=32$). We set $J_{\rm int}=0.1$ and $J_{h}=0.9$.}
\label{Fig9}
\end{figure}

We start to observe the behavior of total density $N_{tot}$ obtained by summing over local density of all sites as varying $\mu$. 
The result is shown in Figs.~\ref{Fig9} (a) and \ref{Fig9} (b). 
We observe how the behavior of the total density of the periodic case as shown in Fig.~\ref{Fig7} changes by introducing edges. 
As same to the periodic boundary calculations in Fig.~\ref{Fig7}, we find some plateaus with the total density constant.
For each plateaus, the bulk states correspond to the bulk SPT phases with different $Z_4$ Berry phase (the density distribution will be shown in later). 

Interestingly, we find two more small plateau regime around $\mu=-0.45$, where particle distribution on edges are specific as shown in later.

\begin{figure}[t]
\begin{center} 
\includegraphics[width=8.5cm]{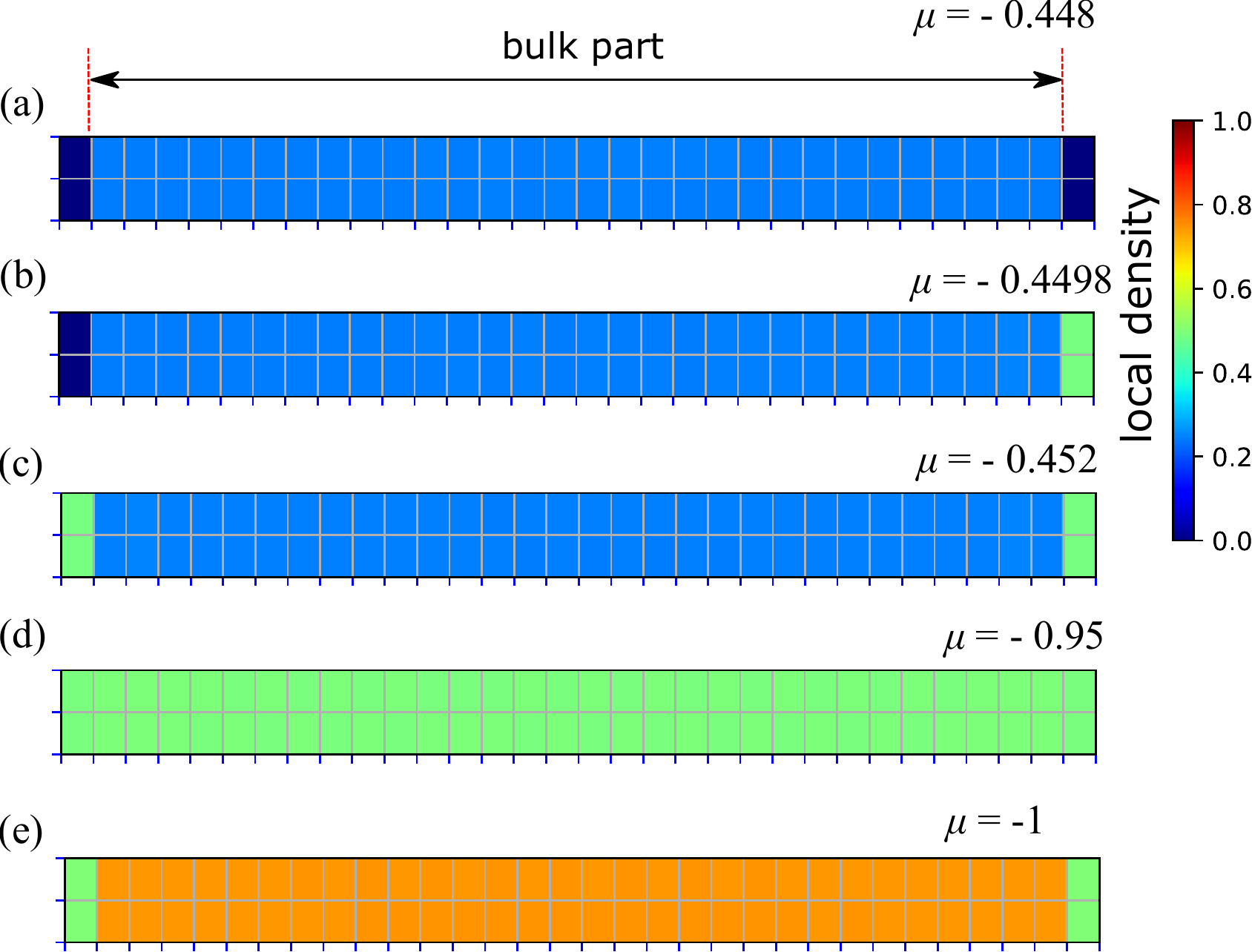}  
\end{center} 
\caption{Density distribution under the diagonal edge for $\mu=-0.448$[(a)], $-0.44498$[(b)], $-0.452$[(c)], $-0.95$[(d)] and $-1$[(e)]. In the data of (d), the edge state is created on the left and right edge sites. The bulk part includes $2L-1$ sites (here we set $L=32$).}
\label{Fig10}
\end{figure}

Let us observe the local density distribution for specific $\mu$'s on each plateaus and what type of edge state appears.
The distribution at $\mu=-0.448$ in the rightmost small plateau in Fig.~\ref{Fig9} (b) is shown in Fig.~\ref{Fig10} (a). 
The bulk part has $\bar{n}=0.25$, corresponding to the phase with $\gamma_\alpha/2\pi=1/4$ and 
the left and right edge sites shows empty (localized holes). Any bosonic edge state does not appear.
Next, we focus on $\mu=-0.4498$ case, this point is in the second rightmost (very) small plateau in Fig.~\ref{Fig9} (b). As shown in Fig.~\ref{Fig10} (b), we here find that a localized edge state at right edge sites appears. 
This state is close to a single bonding state forming on the rung of a ladder \cite{sim_bonding},  $\frac{1}{\sqrt{2}}[b^{\dagger}_{1}+b^{\dagger}_{2}]$. The bulk SPT phase with $\gamma_\alpha/2\pi=1/4$ remains. 
That is, the whole of the wave function can be written by $|\psi^R(\mu=-0.4498)\rangle\sim \frac{1}{\sqrt{2}}[b^{\dagger}_{1,R}+b^{\dagger}_{2,R}]|{\rm bulk}\rangle$, where $|{\rm bulk}\rangle$ is the bulk SPT state.
This edge state is expected from the decoupled plaquette limit for ${\bar n}=0.25$. If we cut the single plaquette into two halves, some bonding-like states can appear. Also we expect that at $\mu=-0.4498$, the ground state is two-fold degenerate. That is, the state  $|\psi^L(\mu=-0.4498)\rangle \sim \frac{1}{\sqrt{2}}[b^{\dagger}_{3,L}+b^{\dagger}_{4,L}]|{\rm bulk}\rangle$ is also another ground state.
Here practically our DMRG calculation just choices the right edge state $|\psi^R(\mu=-0.4498)\rangle$. 
In fact, note that due to the presence of degenerate states, the true observed local density is obtained by $\langle n_j\rangle=\rm{Tr}[\rho_{gs}n_{j}]$ 
where $\rho_{gs}=\frac{1}{2}(|\psi^R(\mu=-0.4498)\rangle \langle\psi^R(\mu=-0.4498)|+|\psi^L(\mu=-0.4498)\rangle \langle\psi^L(\mu=-0.4498)|)$. Then, the densities of left and right edge sites are all same, $0.25$.


We next show the distribution at $\mu=-0.452$ in the third rightmost plateau in Fig.~\ref{Fig9} (b) is shown in Fig.~\ref{Fig10} (c). There, at both edges the localized edge states appear. Varying $\mu$ induces the additional appearance of the edge state. The bulk SPT remains. 
The state is $|\psi(\mu=-0.452)\rangle \sim \frac{1}{2}[b^{\dagger}_{3,L}+b^{\dagger}_{4,L}][b^{\dagger}_{1,R}+b^{\dagger}_{2,R}]|{\rm bulk}\rangle$. 

We further show smaller $\mu$ case. 
The distribution at $\mu=-0.495$ in the leftmost plateau in Fig.~\ref{Fig9} (b) is shown in Fig.~\ref{Fig10} (d). 
The bulk state changes to $\bar{n}=0.5$, corresponding to the SPT phase with $\gamma_\alpha/2\pi=2/4$ and but remains the edge states at both edges. 
The state is $|\psi(\mu=-0.452)\rangle \sim \frac{1}{2}[b^{\dagger}_{3,L}+b^{\dagger}_{4,L}][b^{\dagger}_{1,R}+b^{\dagger}_{2,R}]|{\rm bulk}\rangle$, where $|{\rm bulk}\rangle$ is the SPT with $\bar{n}=0.5$. This numerical result indicates that for the change of $\mu$ around the plateau jumps, $\mu \sim -0.48$, the bulk state changes rather than edge state changes. We expect that the reason is large density fluctuation due to the boson nature. 
It is difficult to find some types of edge states expected from the decoupled plaquett limit for ${\bar n}=0.5$, compared to the low-density region.  

Finally, we observe the distribution at $\mu=-1$ in the left plateau in Fig.~\ref{Fig9} (a) is shown in Fig.~\ref{Fig10} (e).
The bulk part has $\bar{n}=0.75$, corresponding to the phase with $\gamma_\alpha/2\pi=3/4$. 
Same to the previous case as shown in Fig.~\ref{Fig10} (d), the single edge state remains at both edges. 
The state is $|\psi(\mu=-1)\rangle \sim \frac{1}{2}[b^{\dagger}_{3,L}+b^{\dagger}_{4,L}][b^{\dagger}_{1,R}+b^{\dagger}_{2,R}]|{\rm bulk}\rangle$, where $|{\rm bulk}\rangle$ is the SPT with $\bar{n}=0.75$. From this result, the bulk state tends to change rather than edge state around the plateau jumps ($\mu \sim -0.98$).

Summarizing the results of the vertical edge case, we observe that the appearance of the edge state close to the bonding state, identified by difference of density from that of the bulk (mean density ${\bar n}$),  depends on the $Z_4$ SPT phase in the bulk with different mean density ${\bar n}$. In particular, for the bulk SPT with $\gamma_\alpha/2\pi=1/4$, some types of edge state appears by fine-tuning $\mu$. These states are predicted from the state of the decoupled plaquette limit.

\subsection{Diagonal edge case}
\begin{figure}[t]
\begin{center} 
\includegraphics[width=8cm]{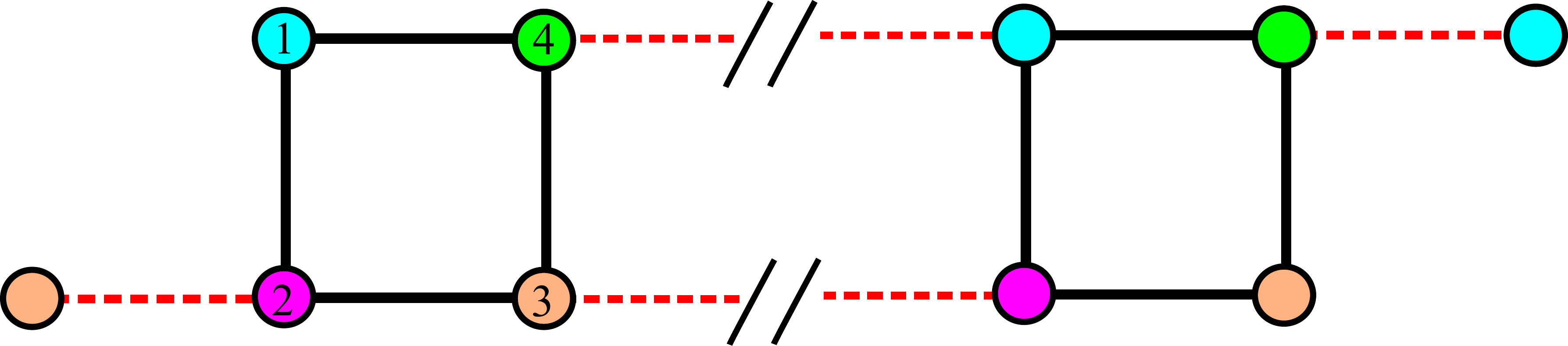}  
\end{center} 
\caption{Schematic image of a diagonal edge. We set $|J_h|\gg |J_{\rm int}|$. 
This open boundary system is invariant under $U_xU_y$ and not invariant under the individual $U_x$ and $U_y$. 
The total lattice site of the system is $L_{tot}=2L+2$.}
\label{Fig11}
\end{figure}

We move to the study of the diagonal edge case. 
The shape of edges is shown in Fig.~\ref{Fig11}. 
Note that this edge geometry cannot be obtained by a periodic ladder ring as in the vertical edge as shown in Fig.~\ref{Fig8}, 
but is obtained by cutting an infinite ladder. 
Here the length of the bulk part is $L$, the total number of the lattice site is $L_{tot}=2L+2$.  
In contrast to the vertical edge case, the boundary condition breaks individual $U_x$ and $U_y$ symmetries, but $U_x \times U_y$  symmetry remains.
We focus on the density distribution for each single sites for the case $|J_{h}|>|J_{\rm int}|$ and numerically investigate whether some edge states appear. Under the diagonal edge, the sites on the edges do not couple to another site in the vertical direction, that is, no vertical hopping and interactions, implying that a localized particle edge state or hole around the edges can appear. 

\begin{figure}[t]
\begin{center} 
\includegraphics[width=7cm]{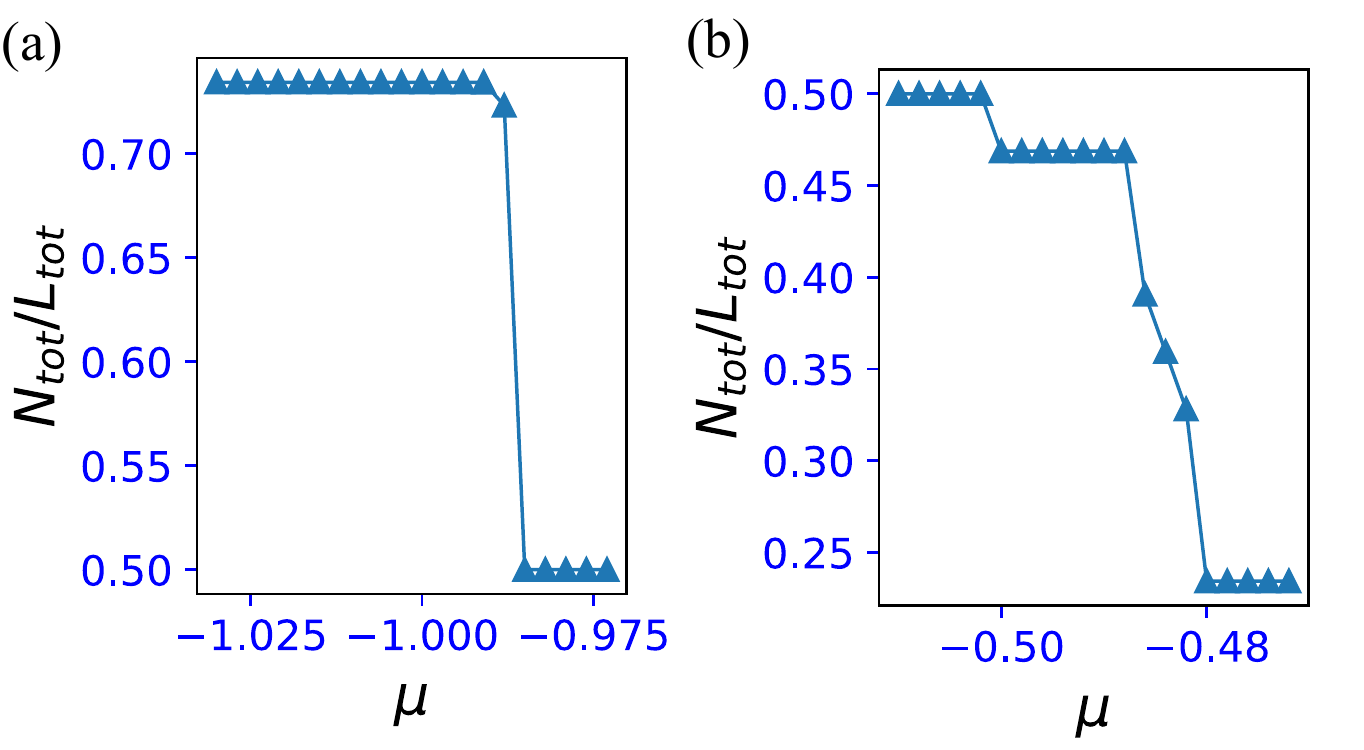}  
\end{center} 
\caption{$\mu$-dependence of total density, $N_{tot}/L_{tot}$.
$L_{tot}=62$ ($L=30$). We set $J_{\rm int}=0.1$ and $J_{h}=0.9$.}
\label{Fig12}
\end{figure}

We start to observe the behavior of total density $N_{tot}$ obtained by summing over local density of all sites as varying $\mu$. 
The result is shown in Fig.~\ref{Fig12} (a) and \ref{Fig12} (b). 
As same to the calculations of periodic boundary condition in Fig.~\ref{Fig7}, we find some plateaus with the total density constant. 
For each plateaus the bulk states correspond to the bulk SPT phases with different $Z_4$ Berry phase. Interestingly, we find one more plateau around $\mu=-0.49$, where particle distribution on edges are specific as shown in later.

\begin{figure}[t]
\begin{center} 
\includegraphics[width=8.5cm]{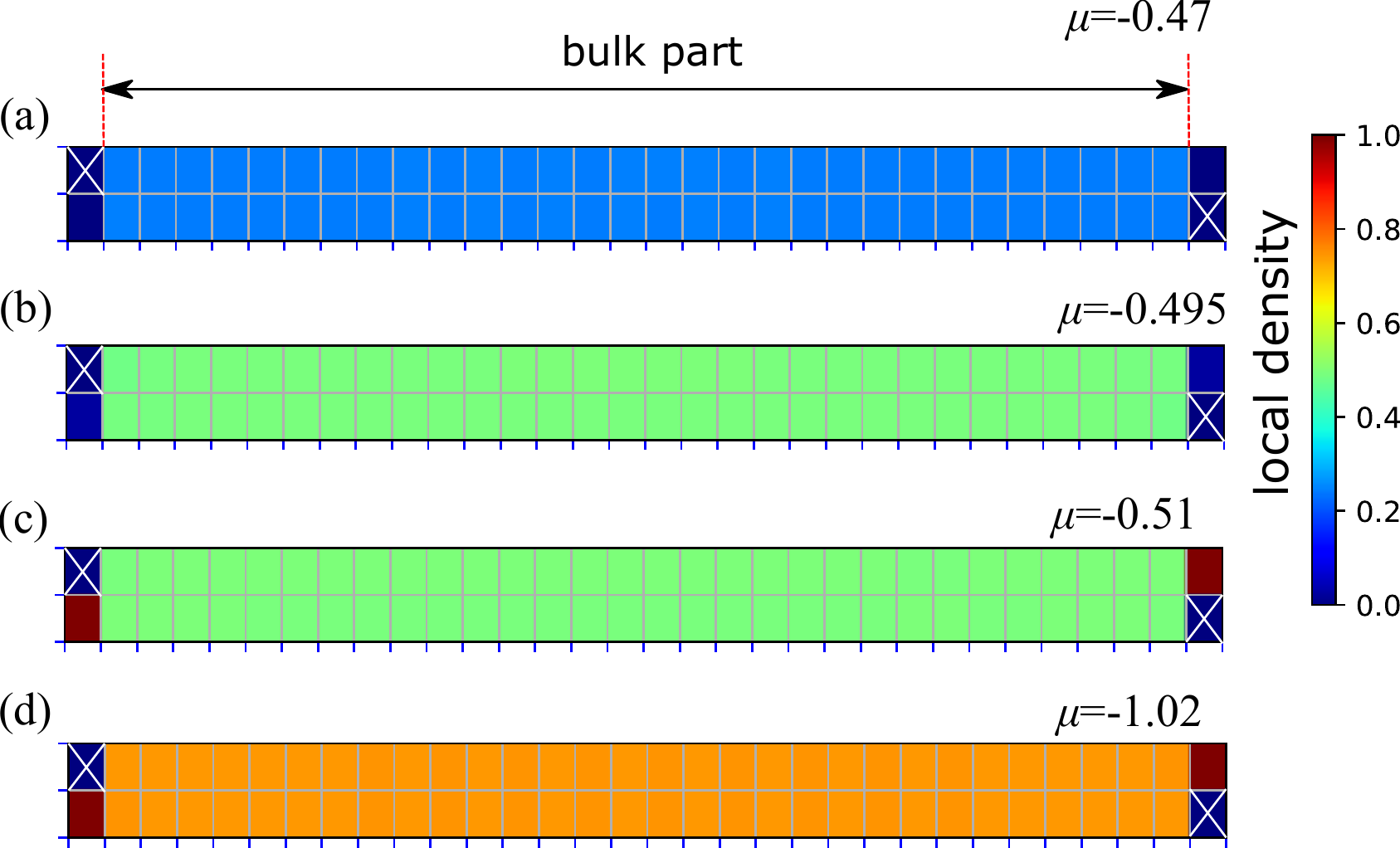}  
\end{center} 
\caption{Density distribution under the diagonal edge for $\mu=-0.47$[(a)], $-0.495$[(b)], $-0.51$[(c)] and $-1.02$[(d)]. The white cross represents a site where the particle occupation is not prohibited due to the diagonal edges. The bulk part includes $2L$ sites (here we set $L=30$).
}
\label{Fig13}
\end{figure}

We further show some local density distributions for specific $\mu$'s on each plateaus.
The distribution at $\mu=-0.47$ in the rightmost plateau in Fig.~\ref{Fig12} (b) is shown in Fig.~\ref{Fig13} (a). 
The bulk part has $\bar{n}=0.25$, corresponding to the phase with $\gamma_\alpha/2\pi=1/4$ and single hole appears at the left and right edge sites. A localized particle edge state does not appear. The reason for it is that if such a particle exists for finite $J_{\rm int}$, the particle tends to intrude to the bulk part due to the low density, but once the particle enters into the bulk (This is likely to occur due to the low density of the bulk), the state is energetically unstable due to the presence of interactions.

Next we show the distribution at $\mu=-0.495$ in the middle plateau in Fig.~\ref{Fig12} (b) is shown in Fig.~\ref{Fig13} (b). 
The bulk part has $\bar{n}=0.5$, corresponding to the phase with $\gamma_\alpha/2\pi=2/4$ and we observe single hole at the left and right edge sites. However, we further a little decrease $\mu$, this plateau is swept and another plateau appears. See the distribution at $\mu=-0.51$ shown in Fig.~\ref{Fig13} (c). Although the bulk SPT state remains same to the case of Fig.~\ref{Fig13} (b). Interestingly a localized particle edge state appears at left and right edge sites, which are much localized at edge sites. Further, note that the localized particle edge state can be a gapless localized particle since the gap between the leftmost and middle plateaus in Fig.~\ref{Fig13} (b) is $1/L$ and for $L\to \infty$, the gap is closed. 

Finally, we observe the distribution at $\mu=-1.02$ in the left plateau in Fig.~\ref{Fig12} (a) is shown in Fig.~\ref{Fig13} (d).
The bulk part has $\bar{n}=0.75$, corresponding to the phase with $\gamma_\alpha/2\pi=3/4$ and interestingly a localized particle edge state appears at left and right edge sites. Any hole at edges does not appear between the rightmost and middle plateaus as shown in Fig.~\ref{Fig13} (b). 

Summarizing the results of the diagonal edge case, some particle edge state much localized on the single edge site appears by fine-tuning $\mu$. 
In general, the presence of interactions exhibits unconventional behaviors of the particle density around the edges. However, on some stable plateaus, we numerically find the presence of clear localized particle edge states, where the local density in the bulk takes a specific constant related to the value of the $Z_4$ Berry phase. 

The presence of the localized particle edge state can be identified as a higher value of density than that of the bulk. 

Finally we comment that 
the local density distributions including edge sites for both vertical and diagonal edge cases presented in this section are observable in real experiments since a recent experimental technique called quantum gas microscope can take a snapshot of local density in an optical lattice and has already identified the presence of edge states from the density profile \cite{Sompet2021}.

\section{Conclusion}
We have proposed a concrete example of interacting SPT phases defined by $Z_2\times Z_2$ symmetry in a Bose Hubbard model on a two-leg ladder. 
The system considered in this work is feasible for a real experiments such as coldatoms in an optical lattice. 
We showed that the $Z_2\times Z_2$ symmetry coming from the lattice geometry leads to a fractional quantization of the $Z_4$ Berry phase. 
The $Z_4$ Berry phase acts as an efficient topological order parameter for the interacting bosonic system.
We numerically demonstrated the presence of the bosonic bulk $Z_4$ SPT phases characterized 
by the fractional quantization of the $Z_4$ Berry phase. 
The phase structure is rich depending on the boson density and strength of interaction, etc. 
Based on the expectation of the presence of the bulk-edge correspondence, we showed case-by-case study to investigate whether or not some edge states appear for two cases by using the DMRG calculation allowing the change of particle number (grand canonical): (i) vertical edge case and (ii) diagonal edge case. 
In the vertical edge case, 
we observed the appearance of the edge state close to the bonding state by fine-tuning $\mu$. In particular, for the bulk SPT with $\gamma_\alpha/2\pi=1/4$, some types of edge state appears by fine-tuning $\mu$. These states are predicted from the state of the decoupled plaquette limit. 
In the diagonal edge case, 
some particle edge state much localized on the single edge site appears by fine-tuning $\mu$, where the bulk states is, of course, SPT.

Finally we comment that it is interesting to extend the ladder geometry to a two-dimensional lattice 
and to investigate whether some higher order topological phase \cite{Benalcazar2017,Araki2020,You2020,Daniel2022} exists based on this $Z_2\times Z_2$ type symmetry as future work.

\bigskip
\section*{ACKNOWLEDGMENTS}
The work is supported by JSPS
KAKEN-HI Grant Number JP21K13849 (Y.K.), 
23K13026 (Y.K.), 23H01091 (Y.H.) and 
JST CREST, Grant No. JPMJCR19T1 (Y.H.), Japan.

\appendix
\renewcommand{\theequation}{A.\arabic{equation} }
\setcounter{equation}{0}
\begin{figure}[t]
\begin{center} 
\includegraphics[width=8.5cm]{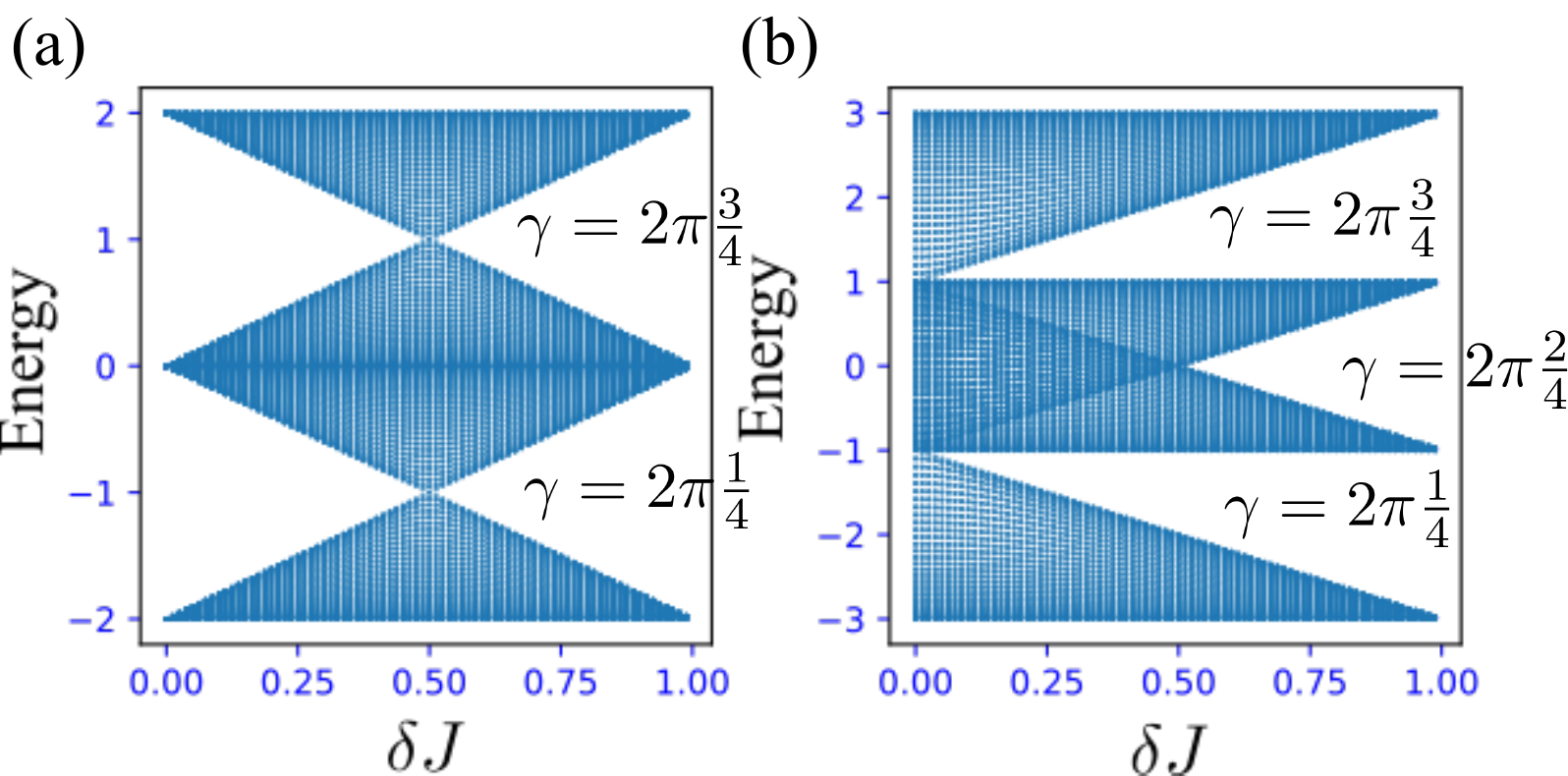}  
\end{center} 
\caption{Single particle spectrum. 
(a) $J_{h}=\delta J$ and $J_{\rm int}=1-\delta J$. 
(b) $J_{h}=1-\delta J$ and $J_{\rm int}=1+\delta J$. 
For both cases, we set $J_v=1$.
}
\label{Fig14}
\end{figure}
\section*{Free spinless fermion system}
The SPT phase protected by the $Z_2\times Z_2$ symmetry can be defined in a free-(spinless)fermion system. 

We calculate the single-particle spectrum of the system for two different parameter sets of $J_1$ and $J_2$ 
with periodic boundary condition. 
The results are shown in Fig.~\ref{Fig14}. 
In Fig.~\ref{Fig11} (a) where we set $J_{\rm int}=1-\delta J$ and $J_h=\delta J$, at $\delta J=0.5$ the first and second gap is closed and reopened for $\delta J>0.5$. 
When the system's fermi-energy resides in the first gap, 
the filling is $1/4$ and there we observe $Z_4$ Berry phase $\gamma/2\pi=\frac{1}{4}$ 
while with the fermi-energy in the second gap, the filling is $3/4$ 
and there we observe $Z_4$ Berry phase $\gamma/2\pi=\frac{3}{4}$. 

We further show another different parameter set where $J_{\rm int}=1-\delta J$ and $J_{h}=1+\delta J$. 
The spectrum is shown in Fig.~\ref{Fig11} (b). 
We observe that the first and third gap appear for $\delta J>0$ while the second gap opens at $\delta J=0.5$. 
With the fermi-energy in the first gap, the filling is $1/4$ and there we observe $Z_4$ Berry phase $\gamma/2\pi=\frac{1}{4}$, with the fermi-energy in the second gap, the filling is $2/4$, there we observe $Z_4$ Berry phase $\gamma/2\pi=\frac{2}{4}$ and 
with the fermi-energy in the second gap, the filling is $3/4$, there we observed $Z_4$ Berry phase $\gamma/2\pi=\frac{3}{4}$.

From these results, the free-fermion system also exhibits the various SPT phase protected by the $Z_2\times Z_2$ symmetry.





\begin{thebibliography}{99}
\bibitem{sim_bonding}
Strictly, the edge states is not strict bonding state due to small $J_{\rm int}$, $\frac{1}{\sqrt{2}}[b^{\dagger}+b^{\dagger}]$. However, our numerical result exhibits the edge state is much close to the bonding state. Thus, we call such a close state edge state in this subsection.
\end{thebibliography}
\end{document}